\documentclass[fleqn,usenatbib,useAMS]{mnras}
\usepackage[utf8x]{inputenc}
\usepackage[english]{babel}
\usepackage{times}
\usepackage{amsmath}
\usepackage{amssymb}
\usepackage{latexsym}
\usepackage{dsfont}
\usepackage{enumitem}

\usepackage{graphicx}
\usepackage[dvipsnames]{xcolor}
\definecolor{myblue}{rgb}{0.0,0.40,0.80}
\definecolor{myred}{rgb}{0.86,0,0.17}
\definecolor{mygray}{rgb}{0.3,0.3,0.3}
\definecolor{roberto}{rgb}{0.0, 0.5, 0.0}

\newcommand{\msun}{\text{${\rm M}_{\odot} \,$}}

\newcommand{\sub}[1]{\scriptsize\raisebox{-2pt}{$#1$}}
\newcommand{\subt}[1]{\scriptsize\raisebox{-2pt}{\text{#1}}} % for roman text

\title[Superdense Cluster of Black Holes]{Dynamics of a Superdense Cluster of Black Holes and the Formation of the Galactic SMBH}

\author[P. Chassonnery and R. Capuzzo--Dolcetta]{
P. Chassonnery,$^{1,2}$\thanks{E-mail~: pauline.chassonnery@ens-cachan.fr} 
R. Capuzzo--Dolcetta$^{2}$\thanks{E-mail~: roberto.capuzzodolcetta@uniroma1.it} \\
$^{1}$Dep. of Mathematics, École Normale Supérieure Paris-Saclay, 61 av. du Président Wilson, Cachan, France; \\
$^{2}$Dep. of Physics, Sapienza, Univ. of Rome, P.le A. Moro 5, Rome, Italy.\\
}

\date{Accepted XXX. Received YYY; in original form ZZZ}

\pubyear{2020}

\begin{document}
\label{firstpage}
\pagerange{\pageref{firstpage}--\pageref{lastpage}}
\maketitle

\begin{abstract}
The center of our Galaxy is known to host a massive compact object, Sgr A$^*$, which is commonly considered as a super-massive black hole of $\sim 4\times 10^6 \,\msun$. It is surrounded by a dense and massive nuclear star cluster, with a half mass radius about $5$~pc and a mass larger than $10^{7} \,\msun$. In this paper we studied the evolutionary fate of a very dense cluster of intermediate mass black holes, possible remnants of the dissipative orbital evolution of massive globular cluster hosts. We performed a set of high precision $N$-body simulations taking into account deviations from pure Newtonian gravitational interaction via a Post Newtonian development up to $2.5$ order, which is the one accounting for energy release by gravitational wave emission. The violent dynamics of the system leads to various successive merger events such to grow a single object containing $\sim 25$ per cent of the total cluster mass before partial dispersal of the cluster, and such to generate, in different bursts, a significant quantity of gravitational waves emission. If generalized, the present results suggest a mechanism of mass growth up to the scale of a super massive black hole.
\end{abstract}

\begin{keywords}
Galaxy~: center --
globular clusters~: general --
stars~: black holes -- 
gravitation --
relativistic processes --
simulations~: $N$-body
\end{keywords}

\section{Introduction}

Many galaxies, including our Milky Way, show evidence of the presence of a compact massive object (CMO) in their centers. These CMOs might be massive or even super-massive black holes (SMBHs) or be in the form of very massive and dense star clusters, commonly referred to as nuclear star clusters (NSCs). The actual `direct' evidence of presence of an SMBH has been recently given by the Event Horizon Telescope (EHT) which gave the first `image' of the shadow produced by the event horizon of a black hole of estimated mass of $6.5\times 10^9 \,\msun$ in the center of the giant elliptical galaxy M87 in the Virgo cluster \citep{EHT2019I}.

Unfortunately, so far, the EHT was not able to provide same evidence for the SMBH of about $4.3 \times 10^6 \,\msun$ \citep{Gillessen2009} allegedly present in the center of our Galaxy. This presence has been clearly suggested by the intense X-ray and radio emission and by the striking observation of the very rapid motion of a certain number of stars very close (within the central arcsec) to the Sgr A$^*$ radio source, as ascertained by two international groups, one at MPE in Garching \citep{Gillessen2009,Schartmann2018} and another one at UCLA \citep{Ghez2005,Boehle2016}. These stars, referred to as S-stars, have been studied over a period of time of about 18 years. One of them, S2 in the denomination given by the MPE group, traveling on its highly eccentric orbit, reached its pericenter distance of about $120$~AU at a speed of $\sim 7650$~km~s$^{-1}$ ($2.55$ per cent of the speed of light). Although $120$~AU is a very close approach ($\sim 4$ times the average Neptune's distance to the Sun and twice the Pluto orbit semi-major axis), it is still well apart from the hypothetical SMBH singularity ($\sim 1400$ Schwarzschild's radii of the hypothetical Sgr A$^*$ black hole). According to \citet{Gravitycollab2018}, the observed gravitational redshift $z \sim 6.7 \times 10^{-4}$ confirms the motion in a regime of strong field. Later, \citet{Gcol20} was able to pick another relativistic effect, namely the prograde precession of the S2 orbit pericenter angle. Anyway, it cannot in principle be excluded that this strong gravitational field at $120$~AU from the GC is due to a super-dense cluster of stars or, more likely, of compact objects. 

The evolution of a very dense stellar system is a quite intriguing and non-trivial issue. Pioneering work in such field was done by~\citet{Spi66} and~\citet{Spi67} although in a necessarily approximate scheme due to the poor computer resources at that time. They found that, unless the stellar system has enough angular momentum to inhibit its contraction at a relatively low stellar density, the process of accelerating contraction of its core must lead inevitably to an increasing number of collisions between the stars in the cluster. Consequently, all stellar aggregations with sufficiently low angular momentum would reach a stage in which direct stellar collisions play a dominant role in the further evolution of the system itself. But the situation of a cluster of compact objects (white dwarves, neutron stars, black holes) would be different because no significant physical collisions would occur to release gas which cools down in the environment possibly giving rise to new stars. So, while the initial phase of core contraction and halo expansion should be similar, the following evolutionary phases of a dense system of normal stars and one composed by compact remnants is likely very different. Such a scheme was later deepened by~\citet{Lig78} who gave an approximate theory of evolution toward core collapse of a cluster composed by stars of two different mass. At this latter regard, is worth citing~\citet{Beg78}. Their qualitative conclusion is that a system composed solely of compact stellar mass bodies would evolve at constant binding energy until a small fraction of the original mass developed into a relativistic bound core, where the physics is of course different and phenomena like energy release by gravitational waves and subsequent merger phenomena cannot be neglected and require a sophisticated treatment. 

The classical computation of the 2-body collision relaxation time scale~\citep{Spi71} gives for the hypothetical cluster of $400$ IMBHs initially packed in a $0.6$~mpc sphere a value of the fraction of a year. Although this very short time scale suggests that the super-violent evolution of the system would lead through a sudden instability to a probable disgregation of the system, much care is due to that the deduction of the relaxation time scale bases on evaluation of diffusion coefficients in the weak scattering regime and, of course, neglecting any relativistic effects. Both these hypotheses are not realized in the real evolution of a dense system of IMBHs, which seems, so, an interesting theme to investigate.

\citet{Kro20} provides a modern vision of the fate of a compact cluster of stellar size BHs left over by evolution of the stellar population originated in starburst clusters residing in the central region of a galaxy short after its formation. Their main finding, based on a modelization which privileges a global view of various evolutionary ingredients respect to accurate N-body modeling in both Newtonian and Post-Newtonian phases, is that the BH cluster compresses down to a relativistic state (velocity dispersion $\sim 3000$~km~s$^-1$) due to insufficient heating by forming BH-BH binaries. The onset of gravitational wave emission implies a loss of mechanical energy which eventually leads to a runaway formation of an SMBH seed, with a $5$ per cent of mass converted into the seed. \\
A somewhat similar result was obtained, still in a scheme which does not include direct N-body simulations, by~\citet{Ant19} who investigate the BH repeated mergers in a dense star cluster ($\rho \gtrsim 10^5 \,\msun$, $v_{esc}\gtrsim 300$~km~s$^{-1}$, conditions fulfilled by $\sim 10$ per cent of present-day NSCs) eventually leading to a very massive ``remnant'' BH. Although upon different approximations and with different methods of study, both~\citet{Ant19} and~\citet{Kro20} agrees on that binary heating is insufficient, at least in a wide range of conditions, to support a cluster of stellar size BHs against collapse.\\

In the above context, the well known dry-merger scenario for the building up of nuclear star clusters (NSCs)~\citep{Tre75,RCD93,Ant12} suggests that orbitally decayed massive globular clusters have carried to the galactic center a quantity of mass such to grow the NSC of the Milky Way and a quantity of intermediate mass black holes (IMBHs). So, the aim of this paper is the study of the evolutionary fate of a possible super dense cluster, as composed of $400$ intermediate mass black holes (IMBHs) of individual mass $10^4 \,\msun$, initially packed in a sphere well within the S2 pericenter distance. Our work represents a significant step forward after the \cite{kup06} paper which studied the dynamics of a dense cluster of compact objects by mean of a modified version of the \texttt{NBODY6++} code~\citep{aar99,spu99} to allow for post-Newtonian effects up to order $2.5$.

The paper is organized as follows~: in Sect.~\ref{section_framework}, the astrophysical frame and the motivations are explained. In Sect.~\ref{section_modmeth}, we describe our methodological approach and the kind of numerical simulations we performed, while in Sect.~\ref{section_results} we discuss the results. Finally, in Sect.~\ref{section_conclusion} we draw the conclusions.

\section{The Astrophysical Framework}
\label{section_framework}

Massive and sufficiently compact objects can decay orbitally in a stellar environment due to the drag caused by the ``wake'' they form behind them during their motion. This is the well known ``dynamical friction'' (df) phenomenon, whose study was pioneered by \citet{Chandra43}. In particular, it has been convincingly shown that massive globular clusters orbiting a galaxy like the Milky Way might decay in the inner region of the host galaxy whenever their orbits are eccentric enough to pass, during their travel across the galaxy, through regions where the environmental phase-space density, whose proxy is $\rho/\sigma^3$ (with $\rho$ and $\sigma$ the local mass density and velocity dispersion), is high enough to induce a significant deceleration.

Actually, the dynamical friction orbital decay has been considered by various authors as a viable explanation for the formation of the nuclear star clusters present in our and other galaxies. The so called \textit{migratory} scenario consists in the orbital decay of a certain number of massive star clusters, followed by their merger in the central region of the galactic potential well. This scenario has been quantitatively validated by many papers~\citep{Tre75,OBS89,Pesce92,RCD93,Cav05,Ase14a,Ase14b}. Here we assume this scenario, which is alternative and/or complementary to the ``in-situ'' model (see e.g.~\citet{Aga11}), to motivate our choice of initial conditions for our evolutionary model. We do not go here into further details, pointing the attention to the recent review on NSCs by~\citet{Neu20}.

\begin{figure}
\centering
	\includegraphics[width=0.7\columnwidth]{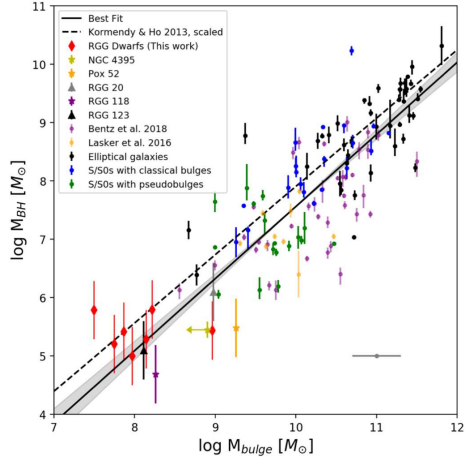}
	\caption{Black hole mass vs host bulge mass (from \citet{sch19}, Figure~4).}
    \label{fig_massBH_massBulge_relation}
\end{figure}

The hypothesis behind our work is that a certain number of massive star clusters (hereafter referred to as globular clusters (GCs)) containing one, or few, intermediate mass black holes (IMBHs) whose mass ranges between few $10^3 \,\msun$ and few $10^4 \,\msun$ have had the time to decay orbitally in an internal region of the host galaxy, carrying with them the hosted IMBHs. The actual presence of such IMBHs, although not clearly confirmed so far by present observations of GCs in the MW halo, would result as a natural interpolation of the host mass vs hosted BH mass correlation over the wide range of scales from open star clusters up to giant elliptical galaxies (see Figure~\ref{fig_massBH_massBulge_relation}).
 
For the BH mass vs host mass,~\citet{sch19} provide (their Eq.~11; see also Fig.˜\ref{fig_massBH_massBulge_relation}) the following fitting formula~:
\begin{equation}
    \textrm{Log}(M_{\subt{BH}}/\msun) = \alpha + \beta \textrm{Log}(M_{\subt{bulge,*}}/(10^{11} \,\msun)), 
\label{fitmbh}
\end{equation}
with $\alpha = 8.80 \pm 0.085$ and $\beta = 1.24 \pm 0.081$.

\begin{figure}
\centering
	\includegraphics[width=0.7\columnwidth]{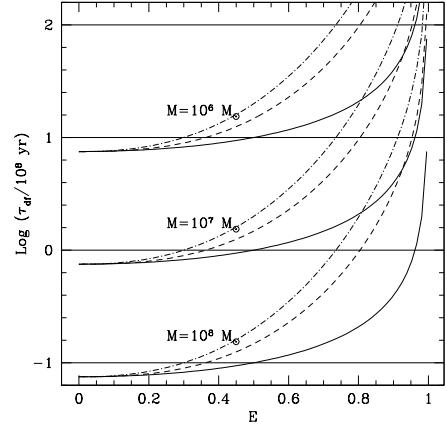}
	\caption{Dynamical friction decay times at varying the object mass $M$ for three values of $J/J_c(E)$ ($0, 0.5, 1$; dot-dashed, dashed and solid line, respectively) in function of the orbital energy $E$. Horizontal lines give the $10^7$, $10^8$, $10^9$ and $10^{10}$~yr thresholds.}
    \label{fig_tdf}
\end{figure}

As we said, an enormous quantity of papers has been dedicated to the topic of the dynamical friction decay time for massive objects, which surely we do not review here, limiting to cite that dynamical friction is, of course, more efficient on massive objects moving on \textit{centrophilic} orbits, that are numerous in non symmetric galactic potentials, whose typical example is the triaxial case. \citet{Pesce92} showed how efficient dynamical friction can be to brake massive clusters in triaxial galaxies, even of moderate axis ratios (1:1.25:2). That work was extended and deepened by~\citet{RCD93} who gave two useful interpolation formulas for the df decay time of a compact cluster moving on both \textit{box} or \textit{loop} orbits in a triaxial potential. Using formulas A1, A2 and A3 of~\citet{RCD93} we computed dynamical friction decay time as functions of orbital energy ($0 \leq E \leq 1$, $E=1$ is the threshold to unbound orbits) and angular momentum scaled to that of circular obit of energy $E$, $J/J_c(E)$. Upon this, we draw Fig.~\ref{fig_tdf}, which shows the dynamical friction times as function of the cluster orbital energy and angular momentum. Almost all GCs with masses larger than $10^7 \,\msun$ would have decayed to the central region of the host galaxy within $1$~Gyr. Additionally, an extrapolation of the fitting formula given by Eq.~\ref{fitmbh} gives, for a hypothetical $10^4 \,\msun$ BH, a host mass of $1.35 \times 10^7 \,\msun$. This means that, if these massive GCs hosted IMBHs at their center, in less than $1$~Gyr they should have carried them to the galactic central region. 

In this frame, we took $10^4 \,\msun$ as individual IMBH mass and decided to study the dynamical evolution of a system of $N_{\subt{BH}} = 400$ black holes, whose summed mass equals, indeed, the estimated Sgr A$^*$ mass, considering them as all initially packed within the innermost pericenter distance of the S stars moving around it, that is $\simeq 0.6$~mpc. This initial configuration is the simplest to adopt, although not the most likely one. Actually, a more reasonable frame would be that where the various GC hosts of the IMBHs shrink their orbit within an assumed galactocentric distance at different times. This frame is more difficult to implement numerically and so its study is postponed to a following paper.

%Formulas A1-A3 in \citet{RCD93} are resumed in
%\begin{equation}
%    \tau_{\subt{df}}(E,J;M) = 10^{g(E,J)} \tau_{\subt{dfb}}(E;M)
%\label{dftime}
%\end{equation}
%where $\tau_{\subt{df}}(E,J;M)$ is the df orbital decay time (in years) of a GC of mass $M$ (in \msun), orbital energy $E$ ($0<E<1$, where $E=1$ is the threshold to unbound orbits) and azimuthal action $J$ ($J=0$ on box orbits). In Eq.~\ref{dftime} the function $g(E,J)>0$ \textit{modulates} the, shortest, df decay time on box orbit ($J=0$) of same energy $E$, $\tau_{\subt{dfb}}(E;M)$. This latter time is given by
%\begin{equation}
%    \tau_{\subt{dfb}}(E;M) = \left(\frac{10^6}{M}\right)\frac{7.5\times 10^8}{\left(1-E\right)^2}\,\textrm{yr},
%\label{tdfb}
%\end{equation}
%with $M$ in \msun, and
%\begin{equation}
%   g(E,J) = 1.7 \, E^{5/2} \sqrt{ {\left( %\frac{J}{J_c} \right)}^{\frac{1}{E+0.1}} },
%\label{gfunc}
%\end{equation}
%where $J_c = J_c(E)$ is the azimuthal action of the quasi-circular orbit of energy $E$. In Figure~\ref{fig_tdf} we show the dynamical friction times as function of $E$ for three values of the GC mass, for each of the three ratios of $J/J_c(E)$ considered. Almost all GC with masses larger than $10^7 \,\msun$ would have decayed to the central region of the host galaxy within $1$~Gyr.

%An extrapolation of the fitting formula given by Eq.~\ref{fitmbh} gives, for a hypothetical $10^4 \,\msun$ BH, a host mass of $1.35 \times 10^7 \,\msun$. This means that, if these massive GCs hosted IMBHs at their center, in less than $1$~Gyr they should have carried them to the galactic central region. 

%In this frame, we took $10^4 \,\msun$ as individual IMBH mass and decided to study the dynamical evolution of a system of $N_{\subt{BH}} = 400$ black holes, whose summed mass equals, indeed, the estimated Sgr A$^*$ mass, considering them as all initially packed within the innermost pericenter distance of the S stars moving around it, that is $\simeq 0.6$~mpc. This initial configuration is the simplest to adopt, although not the most likely one. Actually, a more reasonable frame would be that where the various GC hosts of the IMBHs shrink their orbit within an assumed galactocentric distance at different times. This frame is more difficult to implement numerically and so its study is postponed to a following paper.

\section{Model and Method}
\label{section_modmeth}

The above mentioned very dense cluster of IMBHs is expected extremely prone to instability to collapse, because its $2$-body classical relaxation time scale is of the order (or less than) $10$~yr which is of course a very short time at any respect. Anyway, considerations on the instability of a very dense cluster based on the classical evaluation of the $2$-body relaxation timescale have to be taken with care because, other than that the usual expression of the time scale bases on the unlikely hypothesis that weak $2$-body interaction, they do not account for the possible support against collapse given by binarity. At this regard,~\citet{Kro20} suggest that BH-BH binary heating can be overcome by the huge compression of the BH population inhabiting a massive starburst cluster due to the gas accretion from the environment. So, although the likely fate of our hypothetical superdense cluster is that of a gravitational collapse, its actual modes are not trivial to understand, including the number of mergers and of expelled IMBH, binary fraction and its evolution along the way as well as the possible runaway formation of a SMBH. Recently~\citet{Ant19} deduced a theoretical correlation between the maximum BH mass formed by repeated merger in a dense stellar system and the system characteristics. Dense clusters (density $\gtrsim 10^5 \,\msun$~pc$^{-3}$ and escape velocity $\gtrsim 300$~km s$^{-1}$) lead to BH merger mass up to $10^5 \,\msun$, filling the pair instability strip. This should have relevant counterpart in gravitational wave emission and detection.

Due to the intrinsic non-linearity of the violent dynamical evolution of the cluster, analytical or semi-analytical treatments fail to give precise answers to the many questions that arise, and so we decided to study the evolution of the above mentioned very dense cluster of IMBHs by a direct, high precision, $N$-body approach. The mutual accelerations induced by point-like mass objects packed in a small region of space are so strong that any ``classic'' integration algorithm fails due to the $UV$ divergence of the Newtonian potential. To overcome this problem, we resorted to a high accuracy, regularized code which is our modified version of the algorithmic regularization chain code by Mikkola~\citep{Mikkola2008, Mikkola2010}. The code, called \texttt{ARWV}, and a user manual for it~\citep{cha19}, are freely available to download at \url{https://sites.google.com/uniroma1.it/astrogroup/hpc-html} (the code can be used for scientific publications upon the proper citation condition).\\

The equations of motions of our set of $N$ objects are (for $i=1,2,...,N$)
\begin{equation}
\label{eq_NBsystem}
    \ddot{\mathbf{r}}_{\sub{i}} = G \sum_{\substack{j=1 \\ j\neq i}}^N m_j \frac{\mathbf{r}_{\sub{j}} - \mathbf{r}_{\sub{i}}}{|\mathbf{r}_{\sub{j}} - \mathbf{r}_{\sub{i}}|^3} + \mathbf{f}_{\subt{PN}} + \nabla U_{\subt{ext}} + \mathbf{f}_{\subt{df}}.
\end{equation}

In the formula above, $G$ is the Newton's gravitational constant, $\mathbf{r}_{\sub{i}}$ is the position vector of the generic $i$th object of mass $m_i$, $\mathbf{f}_{\subt{PN}}(\mathbf{r}_{\sub{i}},\mathbf{v}_{\sub{i}})$ is the Post Newtonian (PN) force per unit mass, $\nabla U_{\subt{ext}}(\mathbf{r}_{\sub{i}})$ is the gradient of the external potential, and $\mathbf{f}_{\subt{df}}(\mathbf{r}_{\sub{i}},\mathbf{v}_{\sub{i}})$ is the dynamical friction force per unit mass. 

The Newtonian self interaction is evaluated via a direct summation of all pair contributions in Eq.~\ref{eq_NBsystem}, which implies a computational cost $\mathcal{O}(N^2)$, that limits the use of such kind of high precision codes to a limited number of objects. Another limitation is given, also, by the $UV$ divergence of the Newtonian potential, which makes extremely delicate, on the computational side, dealing with close encounters of massive objects, whose relative acceleration grows enormously at smaller separations. An accurate and elegant, but still computationally expensive, way to deal with those close encounters is via \textit{regularization} of the interaction, which is done by mean of a combination of different techniques (i.e. using (i) logarithmic Hamiltonian~\citep{Mikkola1999a,Mikkola1999b}, (ii) time-transformed leapfrog~\citep{Mikkola2002}, (iii) auxiliary velocity algorithm~\citep{Mikkola2010}). \\

The PN force in Eq.~\ref{eq_NBsystem} is an actual approximation, as expansion in terms of the ratio $(v/c)^2$, to account for general relativistic correction to classic Newton's law of gravitation. The PN approximation was introduced by Einstein, Droste and De Sitter just after publication (in 1916) of the general theory of relativity. The reference paper is~\citet{des16} and a proper summary of PN treatment is found in~\citet{MerrittBook}. Of course, in the limit $(v/c)^2 <<1$ the pure Newtonian interaction is recovered. In $\mathbf{f}_{\subt{PN}}$ we consider PN terms up to the $2.5$ order, i.e. including $\mathcal{O}[(v/c)^2]^{5/2}$ terms, which are the ones needed to account for energy losses via gravitational radiation~\citep{MerrittBook}. We refer to~\citet*{memme04} for the detailed expressions and to~\citet{Mikkola2008} for a description of the actual implementation in the code used here.

Taking into account that GR does not produce $0.5$PN or $1.5$PN contributions to the metric or the equations of motion, the PN force per unit mass acting on the $i$th particle is expressed by
\begin{equation}
    \mathbf{f}_{\subt{PN}}(\mathbf{r}_{\sub{i}},\mathbf{v}_{\sub{i}}) = c^{-2}\mathbf{f}_{\subt{1PN}} + c^{-4}\mathbf{f}_{\subt{2PN}} + c^{-5}\mathbf{f}_{\subt{2.5PN}} + \mathcal{O}(c^{-6}).
\end{equation}

Note that $1$PN and $2$PN terms $(\mathbf{f}_{\subt{1PN}} \mbox{ and } \mathbf{f}_{\subt{2PN}})$ are responsible for pericenter angular shift and are not dissipative (they are symmetric under time reflection $t\rightarrow -t$), while the first dissipative term (\textit{radiation-reaction}) is the $2.5$PN term $(\mathbf{f}_{\subt{2.5PN}})$, which is indeed antisymmetric under time reflection. \\

The $2.5$PN terms (radiation-reaction terms) are responsible for the gravitational wave (GW) emission which extracts mechanical energy from the systems at every merger occurrence. This corresponds to some variation of the mass after merger. We a-posteriori saw that the quantity of energy lost via GW corresponds to a loss of mass $< 0.05$ per cent of the total mass  in all our simulation sets (see Sect.~\ref{sect_grav_waves}), a quantity small enough to justify keeping the mass of individual objects in our simulations unchanged.

Our updated version of \texttt{ARWV} also includes a treatment of an external gravity field in spherical symmetry, due to the presence of a regular distribution of matter in the form of a~\citet{Deh93} and/or a Plummer~\citet{Plummer} profile. A Dehnen (or $\gamma$) density profile is univoquely defined by its total mass $M_{\subt{D}}$, scale radius $r_{\subt{D}}$, and slope parameter $0 \leq \gamma < 3$, while a Plummer profile is characterized by its total mass $M_{\subt{P}}$ and scale radius $r_{\subt{P}}$ only. The role played by the overall, regular, density distribution is that of giving both an additional gravitational acceleration to the point-like objects and a frictional braking, mimicking the cumulative, fluctuating, role of the encounters, via the dynamical friction term, $\mathbf{f}_{\subt{df}}$, in the equations of motion (Eq.~\ref{eq_NBsystem}), which is generally accounted for by mean of the usual Chandrasekhar's expression in local approximation~\citep{Chandra43}~: 
\begin{equation}
\label{eq_dynfric}
    \mathbf{f}_{\subt{df}}(\mathbf{r},\mathbf{v}) = - 4\pi G^2\,\ln\Lambda\,m\, \rho(\mathbf{r}) F(v/\sigma)\, \frac{\mathbf{v}}{v^3}, 
\end{equation}
where $\ln \Lambda$ is the Coulomb logarithm (here assumed $=6.5$), $m$ is the mass of the ``test'' particle, $\rho(\mathbf{r})$ is the local mass density of the field whose $3$D velocity dispersion is $\sigma$, and $\mathbf{v}$ is the velocity of the ``test'' particle. The function $F(v/\sigma)$ is given by
\begin{equation}
    F(v/\sigma)= \textrm{erf}\left( \frac{v/\sigma}{\sqrt{2}} \right) - \sqrt{\frac{2}{\pi}} \frac{v}{\sigma} \textrm{e}^{-\frac{1}{2}(v/\sigma)^2},
\label{eq_dynfric_dfunc}
\end{equation}
where $\textrm{erf}(x)$ is the usual error function, defined as
\begin{equation}
\label{eq_dynfric_erf}
    \textrm{erf}(x) = \frac{2}{\sqrt{\pi}}\int_0^x e^{-t^2}\,dt \leq 1.
\end{equation}

The central environment of the Milky Way can be emulated by a superposition of a Dehnen and a Plummer profile, characterized, respectively, by the sets of values $M_{\subt{D}} = 10^{11} \,\msun$, $r_{\subt{D}} = 2000$~pc, $\gamma_{\subt{D}} = 0.1$ \citep{AS17} and $M_{\subt{P}} = 2.5\times 10^7 \,\msun$, $r_{\subt{P}} = 4$~pc \citep{Sc14}. As it will be shown in Sect.~\ref{section_IC}, for the peculiar initial conditions in study for this article, the actual effects of the external regular distributions of matter (both gravitational acceleration and dynamical friction) result negligible. \\

Let us now give some information about the \texttt{ARWV} code. Usually, after assuming an arbitrary indexing of the $N$ bodies from $1$ to $N$, the position and velocity of each body with respect to the center-of-mass (CoM) of the system are stored in an array of size $6N$. In \texttt{ARWV} the first body, arbitrarily chosen, is considered as a temporary `reference' point and the others bodies are renumbered so as to minimize the distance between the $i$th and $(i+1)$th objects $(i = 1,2,...,N-1)$. With this new numbering the bodies can be seen as forming a `chain' connecting closest to closest body and can be described by their position and velocity, not with respect to the \textbf{CoM} of the system, but with respect to the previous (in term of the chain numbering) body. These `chain' data are stored in an array of size $6(N-1)$ (the first body, being the origin of the chain, is not referenced).

In practice, while creating the chain, the algorithm also tries to minimize the sum of the distance between two successive bodies so as to not inconveniently `forget' any object, that would then have to be added at the end of the chain with an enormous distance to the penultimate object.

The main advantage of this chain scheme resides in that it reduces substantially the round-off errors, making the regularization algorithm more efficient, especially for close interactions between the system bodies. Without this formulation, the step size would reduce to almost zero in critical (very close) encounters. Its downfall is that the interactions are formally much more complicated.

\subsection{Mergers and merger consequences}
\label{section_merger_routine}

During the evolution of an $N$-body system, repeated interactions may lead to the formation of \textit{binaries} (two bodies orbiting one around the other) which may be either temporary or long-living. If long-living, a binary composed by massive objects can eventually merge, losing, first, orbital energy by means of the interaction with the other bodies and, once the binary is tight enough, by means of gravitational radiation which, in our simulations, is accounted for by the $2.5$PN terms. When speaking of massive black holes, this frame is surely important and likely, and needs to be properly accounted for when aiming at a correct simulation of their dynamics. In our \texttt{ARWV} code there is indeed a \textit{merger} routine which enables the code to deal with collisions. The procedure triggers when the distance $r_{ij}$ between two objects of masses $m_{\sub{i}}$ and $m_{\sub{j}}$ is less than $4$ times the sum of their Schwarzschild's radii, that is $r_{\sub{ij}} \leq 8 G (m_{\sub{i}} + m_{\sub{j}})/c^2$. 

To the result of the merger (the \textit{remnant}) is given the location of the center of mass (CoM) of the progenitor pair, though the code halts the integration of the two separate trajectories immediately before that time. For the correct velocity to assign to the remnant (the recoil velocity), there is no consensus. Some authors choose the simplest (but clearly incorrect) choice to give to the remnant a null velocity, while others assume the velocity of the center-of-mass of the two progenitors. In our new version of the \texttt{ARWV} code we introduced a relativistic, spin dependent, recoil velocity, following the prescription given by~\citet{Healy2018}. We shortly describe here the way we did it.

Let $m_{\sub{1}}$ and $m_{\sub{2}}$ be the masses of two merging bodies, with the convention $m_{\sub{1}} \leq m_{\sub{2}}$. Each body is assumed to be spinning; the spin is characterized by a dimensionless spin vector parameter $\boldsymbol{\alpha}_{\sub{i}}$, such that $\alpha_{\sub{i}} \leq 1$. 

Following~\citet{lozl14} and~\citet{Healy2018}, we model the recoil velocity by~:
\begin{equation}
\label{eq_vrec}
    \mathbf{v}_{\subt{rec}} = v_{\sub{m}}\, \mathbf{e}_{\sub{1}} + v_{\sub{\perp}} \left( \cos\xi \,\mathbf{e}_{\sub{1}} + \sin\xi \,\mathbf{e}_{\sub{2}} \right),
\end{equation}
where $\mathbf{e}_{\sub{1}}$ is the unit vector pointing from $m_{\sub{1}}$ to $m_{\sub{2}}$ and $\mathbf{e}_{\sub{2}}$ a unit vector in the orbital plane and orthogonal to $\mathbf{e}_{\sub{1}}$, such that the basis formed by $\mathbf{e}_{\sub{1}}$, $\mathbf{e}_{\sub{2}}$, and the angular momentum of the binary (that is, the mass-weighted sum of the angular momentum vectors of the two progenitor objects) is direct. The quantity $\xi$ is the angle between the ``unequal'' mass contribution to recoil velocity, whose magnitude is $v_{\sub{m}}$, and the spin contribution, of magnitude $v_{\sub{\perp}}$. While both $\xi$ and $v_{\sub{\perp}}$ depend on the values of the spins and of the mass ratio $0 < q = m_{\sub{1}}/m_{\sub{2}} \leq 1$ (see~\citet{Healy2018}), $v_{\sub{m}}$ depends only on $q$, in the following form
\begin{equation}
    v_m = \frac{q^2(q-1)}{(q+1)^5} \left[ A + B\left( \frac{q-1}{q+1} \right)^2 + C\left( \frac{q-1}{q+1} \right)^4 \right],
\end{equation}
where $A = -8712$, $B = -6516$ and $C = 3907$, all in km~s$^{-1}$~\citep{Healy2017,Healy2018}.

\begin{figure}
\begin{tabular}{c}
	\includegraphics[width=0.6\columnwidth,angle=-90]{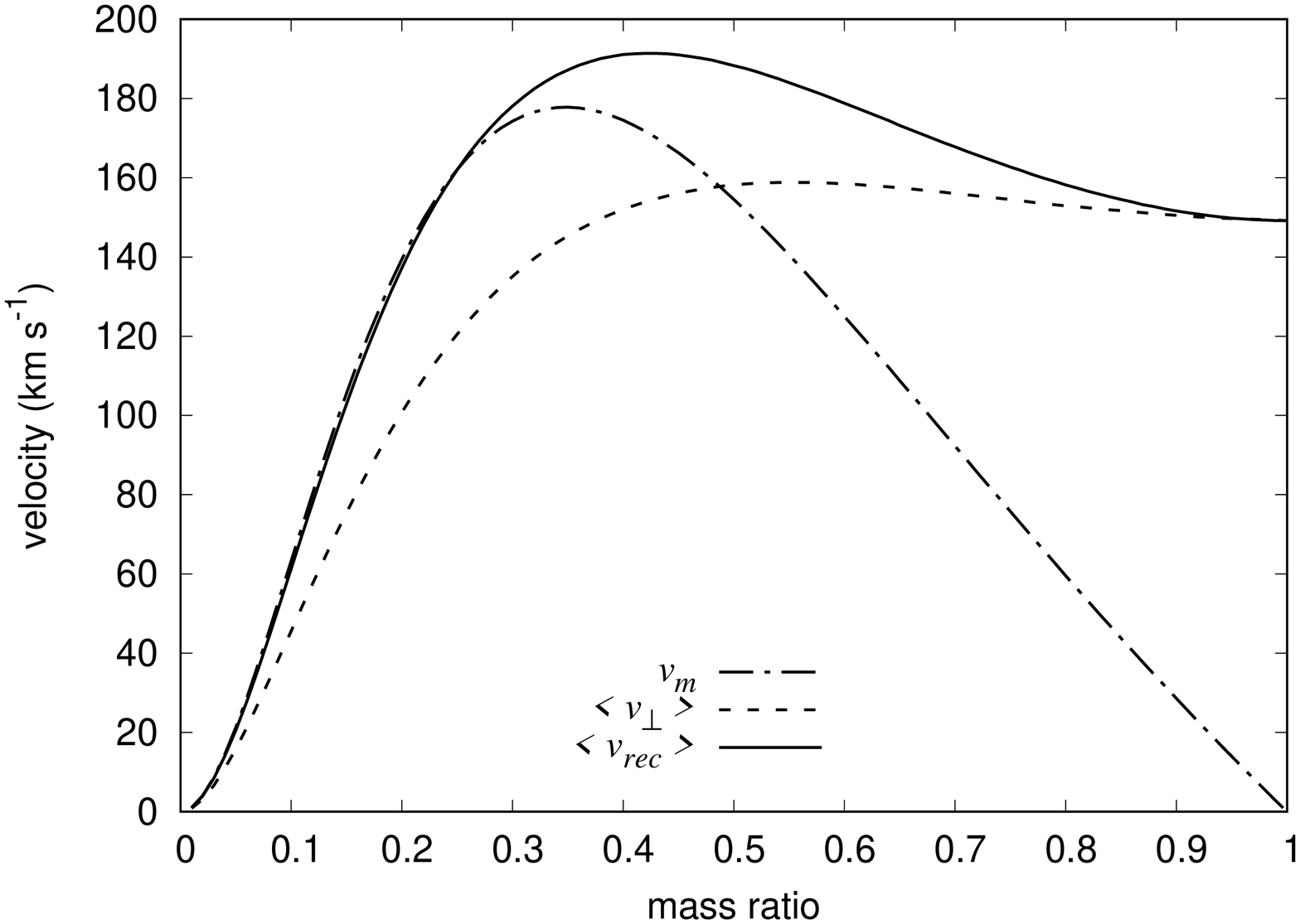} \\
	\includegraphics[width=0.6\columnwidth,angle=-90]{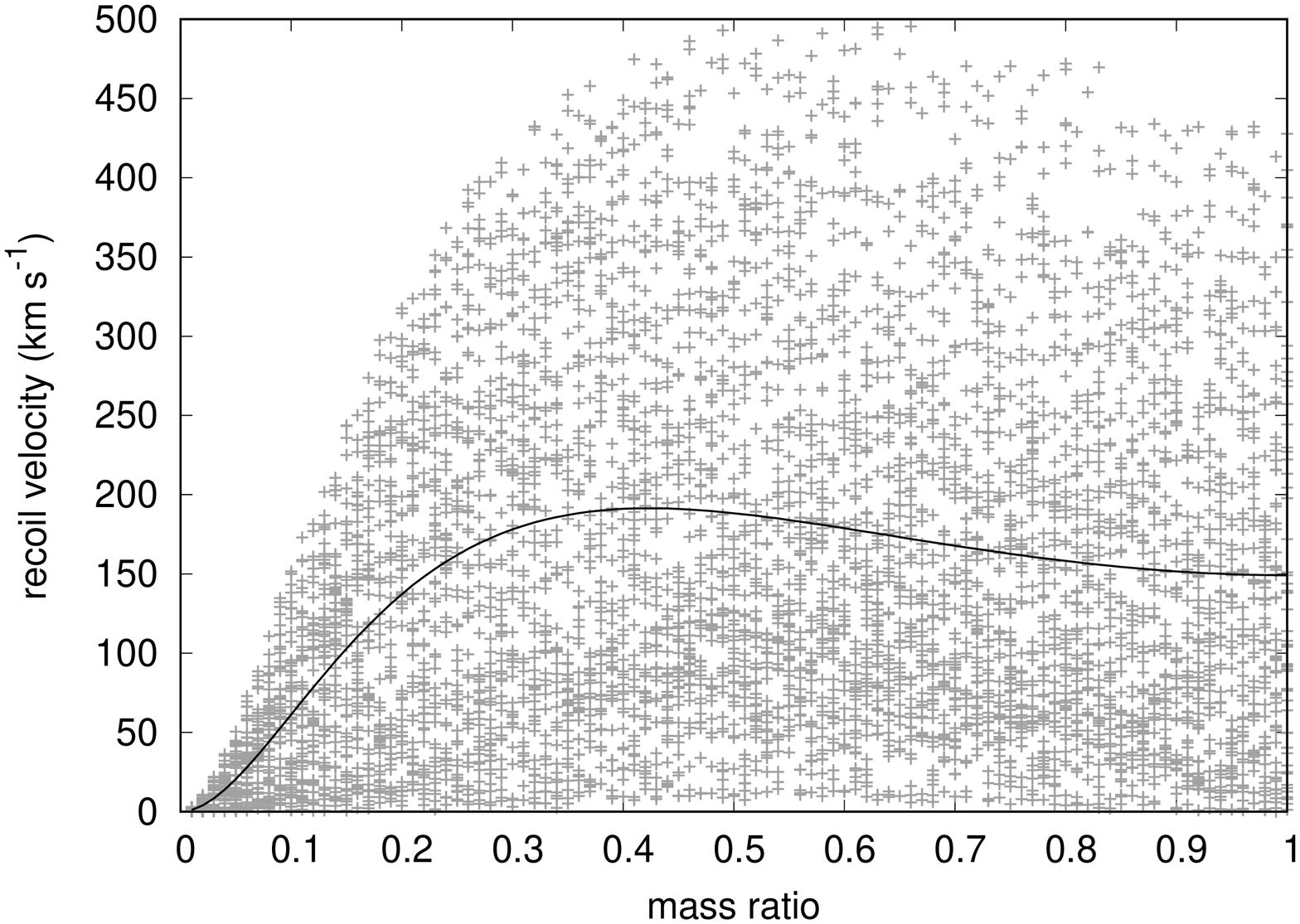}
\end{tabular}
\caption{Top panel~: variation with respect to the mass-ratio of $v_{\sub{m}}$, $v_{\sub{\perp}}$ and $v_{\subt{rec}}$, these two latter averaged over spin (see text). Bottom panel~: distribution of the recoil velocity magnitude $v_{rec}$ (gray crosses) vs mass-ratio, along with its average value (black line).}
\label{fig_recoil_massratio}
\end{figure}

Figure~\ref{fig_recoil_massratio} displays the dependence of $v_{\sub{m}}$, $v_{\sub{\perp}}$ and $v_{\subt{rec}}$ on the mass ratio $q$. Since $v_{\sub{\perp}}$ and $\xi$ (and so $v_{\subt{rec}}$) depend on the dimensionless spins $\boldsymbol{\alpha}_{\sub{1}}$ and $\boldsymbol{\alpha}_{\sub{2}}$, we chose to compute their average values with respect to these parameters. We made a regular sampling of the magnitudes $\alpha_{\sub{1}}$ and $\alpha_{\sub{2}}$ over $[0,1]$ with a step $0.01$, and we took each dimensionless spin as being either ``up'' (i.e. aligned with the angular momentum of the binary) or ``down'' (i.e. antialigned). Then, we averaged the values of $v_{\sub{\perp}}$ and $v_{\subt{rec}}$ obtained for each quadruple $(\alpha_{\sub{1}},\text{up/down}(\boldsymbol{\alpha}_{\sub{1}}),\alpha_{\sub{2}},\text{up/down}(\boldsymbol{\alpha}_{\sub{2}}))$.

By its definition, the angle $\xi$ depends on both the mass ratio and the spins. However, averaging over the uniform spin distribution the dependence upon $q$ is lost, leading to $\langle \xi \rangle_{\sub{\boldsymbol{\alpha}}} = 142.6$°.

We observe that $v_{\sub{m}}$ is maximal for $q \simeq 0.35$, with a nearly linear decrease for $q \geq 0.5$, while $v_{\sub{\perp}}$ is roughly constant for high mass-ratios and so becomes the preponderant part of $v_{\subt{rec}}$, which maximizes at $q \simeq 0.41$. The maximal recoil velocity over all the cases computed was obtained for maximally anti-aligned spins, and it is of the order of $500$~km~s$^{-1}$ (see bottom panel of Figure~\ref{fig_recoil_massratio}). \\

\subsection{Energy variation at merger}
\label{section_envar}

Let us consider a merger between two bodies out of $N$ (arbitrarily numbered $1$ and $2$) happening at time $t_{\sub{m}}$, with $t_{\subt{m}}^{-}$ and $t_{\subt{m}}^{+}$ referring to instant just before and after merger, respectively. For all the bodies save the two undergoing a merger, we have~:
\begin{equation}
    \left\lbrace\begin{array}{l} 
	m_{\sub{i}}(t_{\subt{m}}^{-}) = m_{\sub{i}}(t_{\subt{m}}^{+}), \\
	\mathbf{r}_{\sub{i}}(t_{\subt{m}}^{-}) = \mathbf{r}_{\sub{i}}(t_{\subt{m}}^{+}), \\
	\mathbf{v}_{\sub{i}}(t_{\subt{m}}^{-}) = \mathbf{v}_{\sub{i}}(t_{\subt{m}}^{+}),
\end{array}\right. 
\end{equation}
while for the two merging bodies, the situation is resumed in Table~\ref{tab_merg}. \\

\begin{table}
\centering
\begin{tabular}{ccc}
			& Before merger ($t_{\sub{m}}^{-}$) & After merger ($t_{\sub{m}}^{+}$) \\ \hline \hline
	mass	& $m_{\sub{1}}$ and $m_{\sub{2}}$	& $m_{\subt{rem}} = m_{\sub{1}} + m_{\sub{2}}$ \\ \hline
	position & $\textbf{r}_{\sub{1}}$ and $\textbf{r}_{\sub{2}}$ & $\mathbf{r}_{\subt{rem}} = \mathbf{r}_{\subt{CoM}}$ \\ \hline
	velocity & $\mathbf{v}_{\sub{1}}$ and $\mathbf{v}_{\sub{2}}$ & $\mathbf{v}_{\subt{rem}} = \mathbf{v}_{\subt{CoM}} + \mathbf{v}_{\subt{rec}}$ \\ \hline
	spin & $\boldsymbol{\alpha}_{\sub{1}}$ and $\boldsymbol{\alpha}_{\sub{2}}$ & $\boldsymbol{\alpha}_{\subt{rem}} = \frac{m_1\boldsymbol{\alpha}_{\sub{1}}+m_2\boldsymbol{\alpha}_{\sub{1}}}{m_1+m_2}$ \\ \hline
\end{tabular}
\caption{Parameters characterizing the two (generic) merging objects $m_1$ and $m_2$.}
\label{tab_merg}
\end{table}

Throughout the merger routine, the variation of the kinetic energy, $T$, of the system is~:
\begin{align*}
    \Delta T(t_{\sub{m}}) &= T(t_{\sub{m}}^{+}) - T(t_{\sub{m}}^{-}) \\
    &= \frac{1}{2} \left( m_{\subt{rm}} \mathbf{v}_{\subt{rm}}^2 - m_{\sub{1}} \mathbf{v}_{\sub{1}}^2 - m_{\sub{2}} \mathbf{v}_{\sub{2}}^2 \right) \\
    &= \frac{ m_{\subt{rm}} }{2} \left( v_{\subt{rec}}^2 + 2 \mathbf{v}_{\subt{rec}} \cdot \mathbf{v}_{\subt{CoM}} \right) - \frac{m_{\sub{1}}}{2} \tilde{v}_{\sub{1}}^2 - \frac{m_{\sub{2}}}{2} \tilde{v}_{\sub{2}}^2,
\end{align*}
where the dot $\cdot$ indicates the scalar product, $\tilde{\mathbf{v}}_{\sub{1}} = \mathbf{v}_{\sub{1}} - \mathbf{v}_{\subt{CoM}}$ and $\tilde{\mathbf{v}}_{\sub{2}} = \mathbf{v}_{\sub{2}} - \mathbf{v}_{\subt{CoM}}$. On the other side, the variation of the internal potential energy (that of the pair) is~:
\begin{align*}
    \Delta \Omega_{\subt{int}}(t_{\sub{m}}) &= \Omega_{\subt{int}}(t_{\sub{m}}^{+}) - \Omega_{\subt{int}}(t_{\sub{m}}^{-}), \\
    &= - \sum_{j=3}^{N} \frac{G m_{\subt{rm}} m_{\sub{j}} }{|\mathbf{r}_{\subt{rm}} - \mathbf{r}_{\sub{j}}|} + \frac{G m_{\sub{1}} m_{\sub{2}} }{|\mathbf{r}_{\sub{1}} - \mathbf{r}_{\sub{2}}|} + \sum_{i=1,2} \sum_{j=3}^{N} \frac{G m_{\sub{i}} m_{\sub{j}} }{|\mathbf{r}_{\sub{i}} - \mathbf{r}_{\sub{j}}|}, \\
    &\simeq - \sum_{j=3}^{N} \frac{G m_{\subt{rm}} m_{\sub{j}} }{|\mathbf{r}_{\subt{rm}} - \mathbf{r}_{\sub{j}}|} + \frac{G m_{\sub{1}} m_{\sub{2}} }{|\mathbf{r}_{\sub{1}} - \mathbf{r}_{\sub{2}}|} + \sum_{i=1,2} \sum_{j=3}^{N} \frac{G m_{\sub{i}} m_{\sub{j}} }{|\mathbf{r}_{\subt{rm}} - \mathbf{r}_{\sub{j}}|}, \\
    &\simeq \frac{G m_{\sub{1}} m_{\sub{2}} }{|\mathbf{r}_{\sub{1}} - \mathbf{r}_{\sub{2}}|},
\end{align*}
considering that, for any pair $(i,j)$ with $i=1,2$ and $j \in \{3,...,N\}$, we have $\tilde{r}_{\sub{i}} \equiv |\mathbf{r}_{\sub{i}} - \mathbf{r}_{\subt{CoM}}| \ll |\mathbf{r}_{\subt{CoM}} - \mathbf{r}_{\sub{j}}|$, and so $|\mathbf{r}_{\sub{i}} - \mathbf{r}_{\sub{j}}| \simeq |\mathbf{r}_{\subt{CoM}} - \mathbf{r}_{\sub{j}}| = |\mathbf{r}_{\subt{rm}} - \mathbf{r}_{\sub{j}}|$. Finally, the variation of the external potential energy is~:
\begin{align*}
    \Delta \Omega_{\subt{ext}}(t_{\sub{m}}) &= \Omega_{\subt{ext}}(t_{\sub{m}}^{+}) - \Omega_{\subt{ext}}(t_{\sub{m}}^{-}), \\
    &= - m_{\subt{rm}} U_{\subt{ext}}(\mathbf{r}_{\subt{rm}}) + m_{\sub{1}} U_{\subt{ext}}(\mathbf{r}_{\sub{1}}) + m_{\sub{2}} U_{\subt{ext}}(\mathbf{r}_{\sub{2}}), \\
    &= \mathcal{O}\left( (\tilde{r}_{\sub{1}}^2 + \tilde{r}_{\sub{2}}^2) \sup(U''_{\subt{ext}}) \right).
\end{align*}

Given the above considerations, the total variation of the mechanical energy of the $N$-body system during the merging process is~:
\begin{equation}
    \Delta E(t_{\sub{m}}) \simeq \frac{ m_{\subt{rm}} }{2} v_{\subt{rec}}^2 + m_{\subt{rm}} \mathbf{v}_{\subt{rec}} \cdot \mathbf{v}_{\subt{CoM}} - E_{\sub{b}},
    \label{eq_deltaE_merger}
\end{equation}
with $E_{\sub{b}} = \frac{m_{\sub{1}}}{2} \tilde{v}_{\sub{1}}^2 + \frac{m_{\sub{2}}}{2} \tilde{v}_{\sub{2}}^2 - \frac{G m_{\sub{1}} m_{\sub{2}} }{|\mathbf{r}_{\sub{1}} - \mathbf{r}_{\sub{2}}|}$ the internal energy of the progenitor pair, which is negative in the case of a bound binary. So, neglecting the term $\mathbf{v}_{\subt{rec}} \cdot \mathbf{v}_{\subt{CoM}}$ which, on average over numerous merger events, should be null, we found that $\sum\limits_{i \geq 1} \Delta E(t_{\sub{m}}^i) > 0$.

\subsection{Initial conditions}
\label{section_IC}

As we said above, our aim is that of simulating an extremely dense stellar cluster which could be a precursor of the Milky-Way central SMBH. As total mass of our system we assumed $M_{\sub{S}} = 4 \times 10^{6} \,\msun$, composed by $N = 400$ IMBHs of same individual mass $m = 10^4 \,\msun$.

The initial homogeneous and virialized very packed configuration was motivated by willing to verify how unstable such distribution were. Actually, a possible question could have been~: is it possible that a superdense system of gravitating objects distributed around the Galactic center lives long enough to make the surrounding S-stars moving as they are presently seen without invoking the presence of a black hole singularity ? Answering to this question with a full $N$-body simulation requires, indeed, `packing' the 400 IMBHs in a sphere of initial radius, $R_{\sub{0}}$, sufficiently smaller than the smallest pericenter of the S-stars (note that the innermost pericenters of the S-stars, those of S2 and S14, are about $5$~mpc, where $1$~mpc = $1$~milliparsec = $10^{-3}$~pc).

A first set of simulations (hereafter referred to as set~1) was conducted with $R_{\sub{0}} = 0.6$~mpc, that is $\sim$ ten times less than the smallest S-star pericenter distance. Under these quite extreme conditions ($\rho_{\sub{0}} \sim 4.4 \times 10^{15} \,\msun$~pc$^{-3}$ !), the cluster is expected to undergo a fast dynamical instability, so that, also for the sake of comparison, we run a second set of simulations (called set~2) with a 10 times larger initial radial size, $R_{\sub{0}} = 6$~mpc (same size of closest S-star pericenter distance).
In the hypothesis of uniform spatial distribution, the central escape velocity in set~1 and set~2 is, respectively, $v_{e,1}\simeq 7.11 \times 10^3$~km~s$^{-1}$ and $v_{e,1}\simeq 2.25 \times 10^3$~km~s$^{-1}$.

As we said, the distribution of the initial positions ($\mathbf{r}_{\sub{i}}$, for $i=1,2,...,N$) of the $N = 400$ IMBHs has been assumed uniform within $R_{\sub{0}}$, practically obtained by a standard pick-and-reject method.

The initial velocity distribution ($\mathbf{v}_{\sub{i}}$, for $i=1,2,...,N$) was assumed, also, randomly generated according to a uniform isotropic distribution scaled such as to give a chosen initial virial ratio $Q_{\sub{0}} \equiv 2 T_{\sub{0}} /|\Omega_{\sub{0}}| = 1$ (where $T$ and $\Omega$ are the total kinetic and potential energy, respectively). 
%The potential energy accounts also for the regular distribution of matter~:

%\begin{equation}
%\label{eq_potential_energy}
%\Omega = -\sum_{i=1}^{N} \left[ \sum_{j>i} \frac{Gm^2}{|\mathbf{r}_{\sub{i}} - %\mathbf{r}_{\sub{j}}|} + m U_{\subt{ext}}(\mathbf{r}_{\sub{i}})\right].
%\end{equation}

% Due to the chaoticity of the classical gravitational $N$-body problem, the long term results of a simulation actually vary also in dependence of the specific hardware and software used for the computations, due to loss of control of the round-off error. 

To give some statistical reliability to our dynamical experiments, we have performed a total of $40$ simulations with different sampling of the same initial conditions. Practically, for both set~1 and set~2 we generated $10$ input files using the same global parameters ($N = 400$, $M_{\sub{S}} = 4\times 10^6 \,\msun$, $R_{\sub{0}} = 0.6$~mpc for set~1, and $R_{\sub{0}} = 6$~mpc for set~2) but choosing different random seeds to sample the same (homogeneous) spatial density and velocity distributions. Moreover, we randomly generated one set, \texttt{s1}, of dimensionless spin vectors for the IMBHs ($\boldsymbol{\alpha}_i$, for $i=1,2, \dots, N$) following a uniform distribution in a sphere of unitary radius. For both set~1 and set~2, we then performed a first subset of $10$ simulations, called set~1A and 2A, with all spins equal to zero, and a second subset, named 1B and 2B, of $10$ simulations each, where the spins are selected according to the procedure above. A sketch of the main parameters of the various simulations is given in Table~\ref{tab_param}.

To enhance accuracy in the computations, in the code we use $R_{\sub{0}}$ as length unit and $M_{\subt{tot}} = M_{\sub{S}} + M_g$ as mass unit, with $M_g$ the galactic mass inside the sphere of radius $R_{\sub{0}}$. The time unit $\textrm{U}_{\subt{t}}$ is chosen so as to ensure $G=1$~:
\begin{equation}
\label{eq_timeunit}
    \textrm{U}_{\subt{t}} = \frac{ R_{\sub{0}}^{3/2} }{ \sqrt{G M_{\subt{tot}}} } = \left\lbrace\begin{array}{l}
        0.107 \mbox{ yr, for set } 1, \\
        3.365 \mbox{ yr, for set } 2.
    \end{array}\right.
\end{equation}

The above time is, actually, the typical crossing time of the system.
Due to the huge space density of the IMBH cluster under study, the dynamics is very violent and computationally demanding. Moreover, the computational cost of the planned simulations clearly varies as $\mathcal{O}(t_{\subt{max}}/\textrm{U}_{\subt{t}})$, so that, to integrate up to the same physical time $t_{\subt{max}}$, a simulation of set~1 would require, a priori, a $\sim 30$ times longer (in terms of CPU time) simulation than one of set~2. Of course, many other issues have an impact on the computational speed, and indeed different simulations pertaining to the same set (1 or 2) proceeded at different speed. Therefore, we decided to simulate the evolution of the system over $4000\, \textrm{U}_{\subt{t}}$ in each case, which means that for the denser configurations of set~1 we have $t_{\subt{max}} = 426$~yr while for set~2 we have $t_{\subt{max}} = 13\ 462$~yr (that is a factor $31.6$ in terms of physical time).

\begin{table}
\centering
\begin{tabular}{|c|c|c|c|}
    Set & $R_{\sub{0}}$ (mpc) & $t_{\subt{max}}$ (yr) & spin \\ \hline
    1A & $0.6$ & $426$ & zero \\ \hline
    1B & $0.6$ & $426$ & uniform \\ \hline
    2A & $6$   & $13\ 462$ & zero \\ \hline
    2B & $6$   & $13\ 462$ & uniform \\ \hline
\end{tabular}
\caption{Main parameters of the various sets of simulations. $t_{\subt{max}}$ is the maximum time extension of the simulation.}
\label{tab_param}
\end{table}

\subsection{The actual role of the recoil velocity}

\begin{figure}
\centering
\begin{tabular}{c}
	\includegraphics[width=0.6\columnwidth,angle=-90]{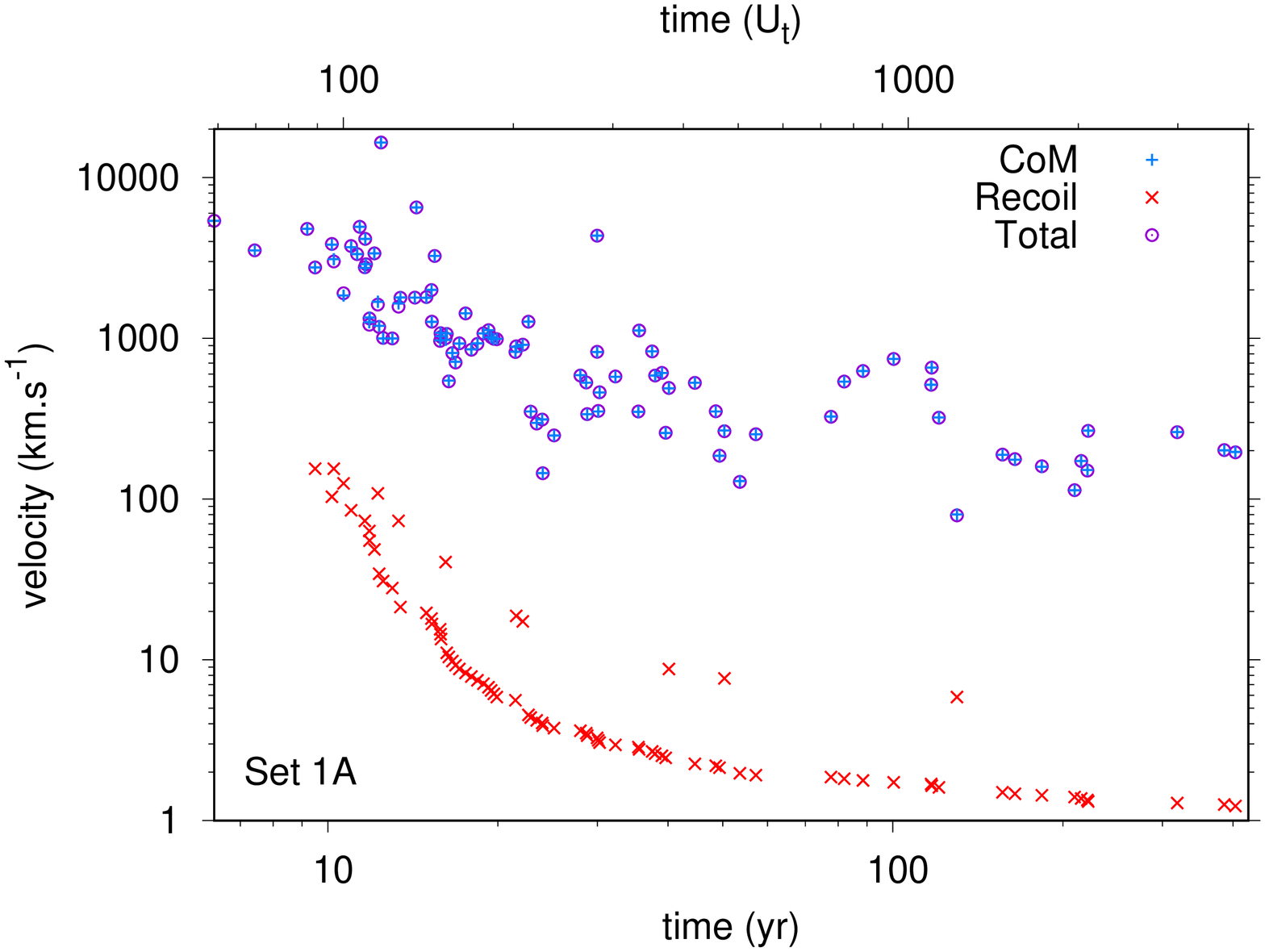} \\
	\includegraphics[width=0.6\columnwidth,angle=-90]{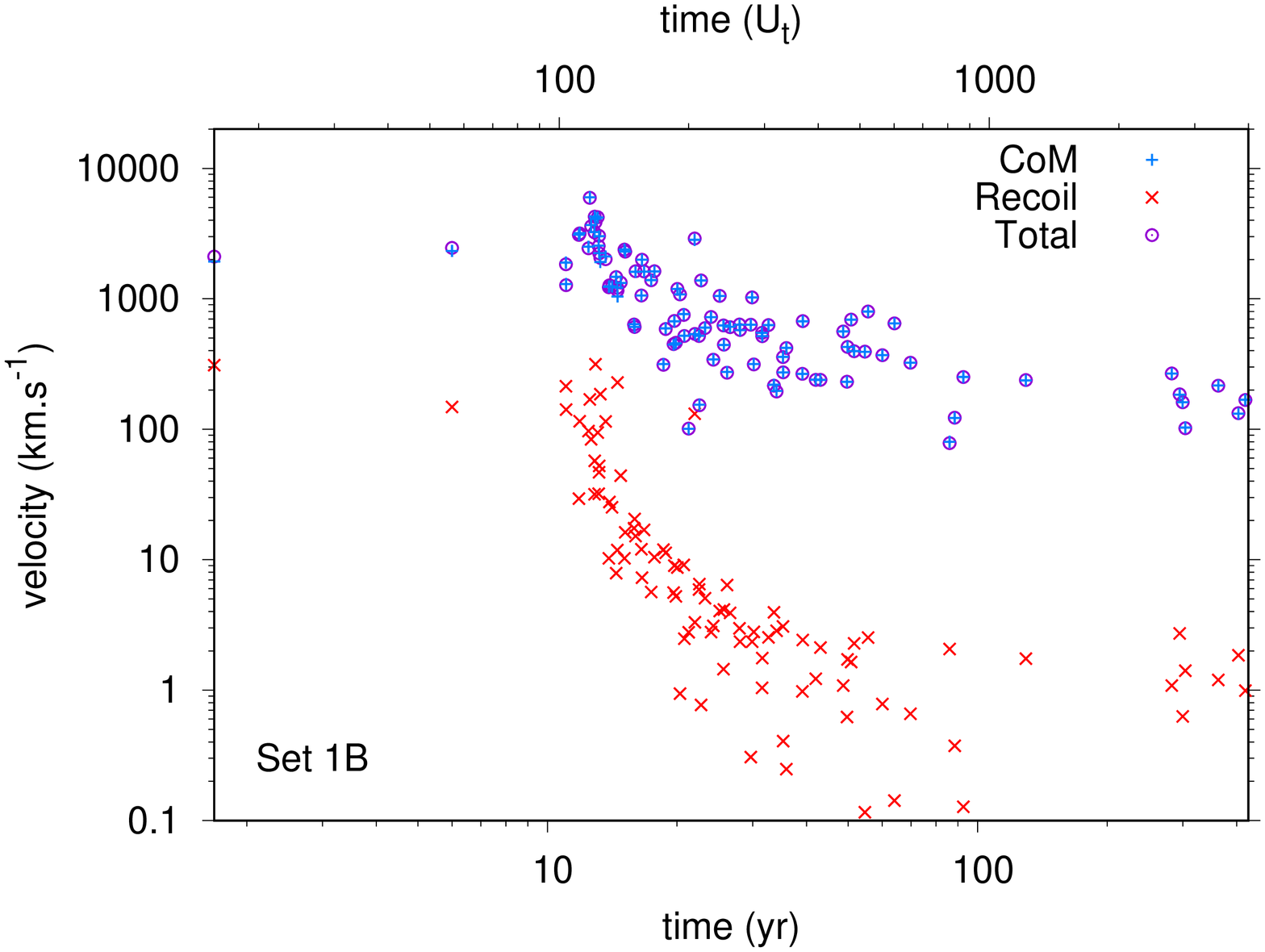} \\ 
	\includegraphics[width=0.6\columnwidth,angle=-90]{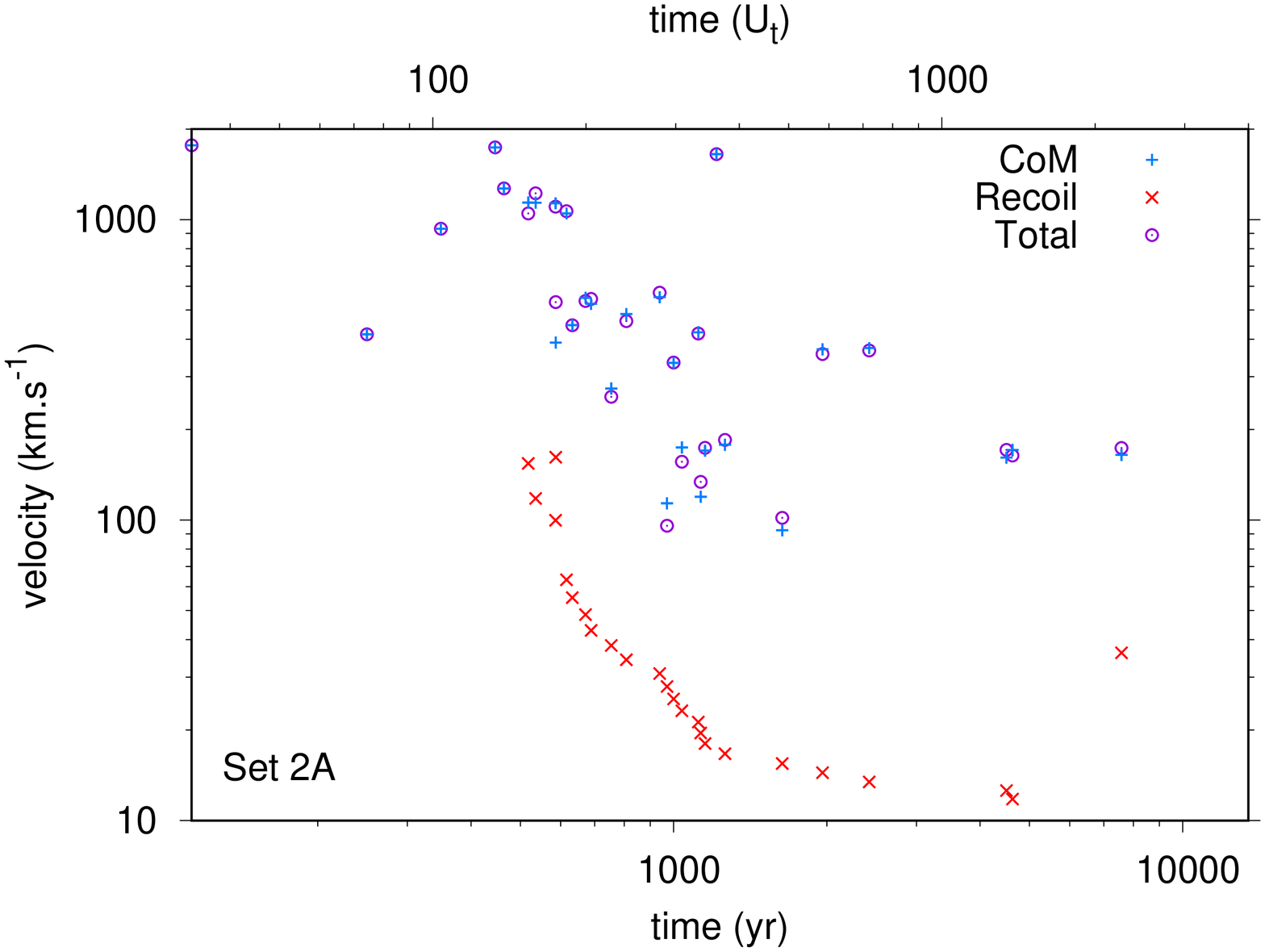} \\
	\includegraphics[width=0.6\columnwidth,angle=-90]{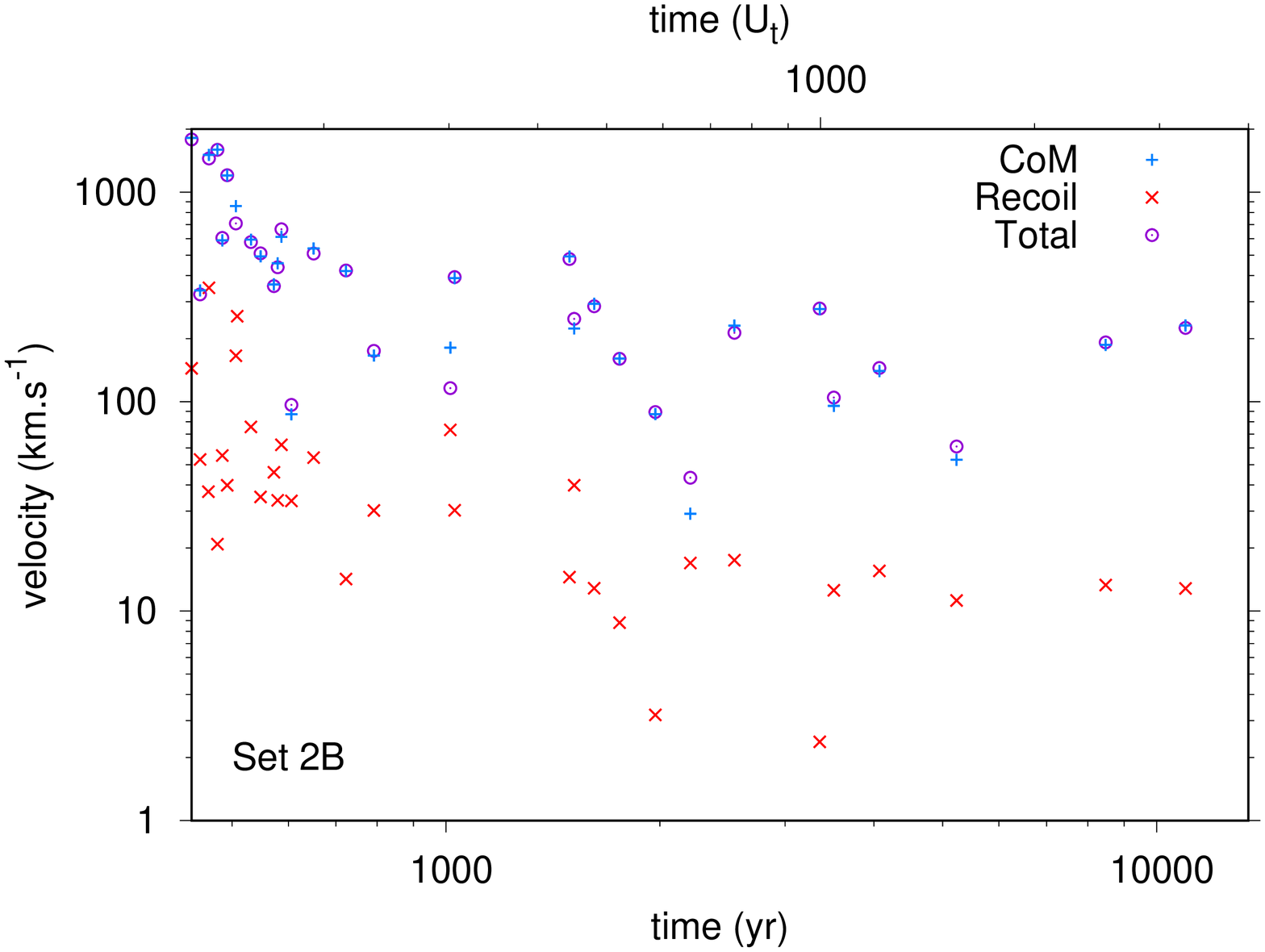} \\
\end{tabular}
\caption{Contribution of the recoil velocity (in red) and progenitor binary center-of-mass velocity (in blue) to the total velocity of the remnant (in purple) at each merger, over one simulation of each subset. From top to bottom~: set~1A, set~1B, set~2A, set~2B.}
\label{fig_recoil_data}
\end{figure}

For set~1, corresponding to the densest cluster, the rescaled initial velocities range from a few hundred km~s$^{-1}$ to $\sim 5400$~km~s$^{-1}$, with an average value $\langle v \rangle = 4012$~km~s$^{-1}$. On the other hand, the recoil velocity after merger is at most of $500$~km~s$^{-1}$ ($200$~km~s$^{-1}$ on average), generally small with respect to the velocity of the center-of-mass of the precursor binary and not large enough to overcome the escape velocity ($\simeq 7110$ km s$^{-1}$).The two top panels of Figure~\ref{fig_recoil_data} (which refer to set~1) suggest that the recoil velocity alters the course of some of the individual trajectories, but it does not have, on average, a very significant impact on the overall evolution of the cluster.

In set~2, where the IMBHs are initially less densely packed, the initial velocities range from $100$~km~s$^{-1}$ to $\sim 1700$~km~s$^{-1}$, with average value $\langle v \rangle = 1270$~km~s$^{-1}$. Due to the lower escape velocity, the recoil velocity is expected to have a more relevant impact on the course of the simulation than in set~1. Anyway, as shown in the two bottom panels of Fig.~\ref{fig_recoil_data}, the recoil velocity is still one order of magnitude smaller than the progenitor binary center-of-mass velocity and of the escape velocity which is $\simeq 2250$ km s$^{-1}$). Therefore, it can rarely cause the ejection of a merger remnant from the main cluster by overcoming the local escape velocity.  Note the decrease with time of both center-of-mass and recoil velocity in both set~1 and set~2, explained by decreasing in time of $q$.

Because the effect of the recoil velocity is mostly negligible, there is no statistical difference neither between the results of the subset~1A and 1B nor between the results of the subset~2A and 2B.

For this reason, in the rest of this work we will, as a rule, only present results averaged over the whole set~1 and the whole set~2, and not detailed subset's results.

\subsection{The actual role of the external potential}
\label{section_extpot}

As we said in Sect.~\ref{section_modmeth}, for our simulations, we emulate the central environment of the Milky Way by an external density profile which is the superposition of a Dehnen profile characterized by $M_{\subt{D}} = 10^{11} \,\msun$, $r_{\subt{D}} = 2000$~pc and $\gamma_{\subt{D}} = 0.1$ \citep{AS17} and a Plummer profile characterized by $M_{\subt{P}} = 10^7 \,\msun$ and $r_{\subt{P}} = 5.4$~pc \citep{Sc14}.

The galactic mass inside the sphere of radius $R_{\sub{0}}$, $M_g(R_{\sub{0}})$, is in both the sets of simulations very small with respect to the total mass in IMBHs ($M_g/ M_{\sub{S}} = 2.4\times 10^{-13}$ for set~1 and $2.4\times 10^{-10}$ for set~2), so that the external field is altogether negligible in terms of gravitational acceleration respect to the pairwise IMBH-IMBH gravitational acceleration. Moreover, for small distances $r$ to the center ($r \sim 1$~mpc), the background density is dominated by the Plummer profile, which is flat for $r \ll r_{\subt{P}} = 5.4$~pc. Hence, for both sets of our simulations, the background density averaged within the sphere of radius $R_{\sub{0}}$ has the same value $\left\langle \rho(<R_{\sub{0}}) \right\rangle \simeq 1.5 \times 10^4 \,\msun$~pc$^{-3}$. We can estimate $\left\langle \sum\limits_{1 \leq i < j \leq N} |\mathbf{r}_{\sub{i}} - \mathbf{r}_{\sub{j}}|^{-2} \right\rangle \simeq 3N/(4R_{\sub{0}}^2)$, so that the initial effect of dynamical friction $\mathbf{f}_{\sub{df}}$ with respect to the Newtonian interactions $\mathbf{f}_{\sub{N}}$ can be quantified as~:
\begin{align}
\label{eq_compare_newton_dynfric}
    \frac{|\mathbf{f}_{\subt{df}}|}{|\mathbf{f}_{N}|} &\simeq 16\, \pi G \ln\Lambda \langle \rho(<R_{\sub{0}})\rangle \left\langle \frac{ F\left( v/\sigma \right) }{v^2} \right\rangle \frac{R_{\sub{0}}^2}{3N}, \\ 
    &\simeq \left\lbrace\begin{array}{l}
        6.5 \times 10^{-13},\, \mbox{ for set } 1, \\
        6.5 \times 10^{-10},\, \mbox{ for set } 2, 
    \end{array}\right. \nonumber
\end{align}
which is totally negligible. Anyway, the role of external potential is relevant to determine the fate of objects that, during the various interactions and also after mergers, acquire a speed sufficient to move far from the center. Most of them do not overcome the escape velocity and so make a fast return to the internal region due to the combined action (gravitational acceleration and dynamical friction) of the external field. This slows down the cluster dissolution.

\section{Results}
\label{section_results}

Here we present results for our sets of simulations, whose characteristics have been described in Sect.~\ref{section_IC} and summarized in Table~\ref{tab_param}. Results are indicative on the overall fate of the super dense cluster of IMBHs and show the clear growth of a super-massive black hole seed via subsequent merger events, each of them characterized by a burst of gravitational wave emission.

\subsection{Overall evolution of the cluster}

\begin{figure}
\centering
\begin{tabular}{r}
	\includegraphics[width=0.71\columnwidth,angle=-90]{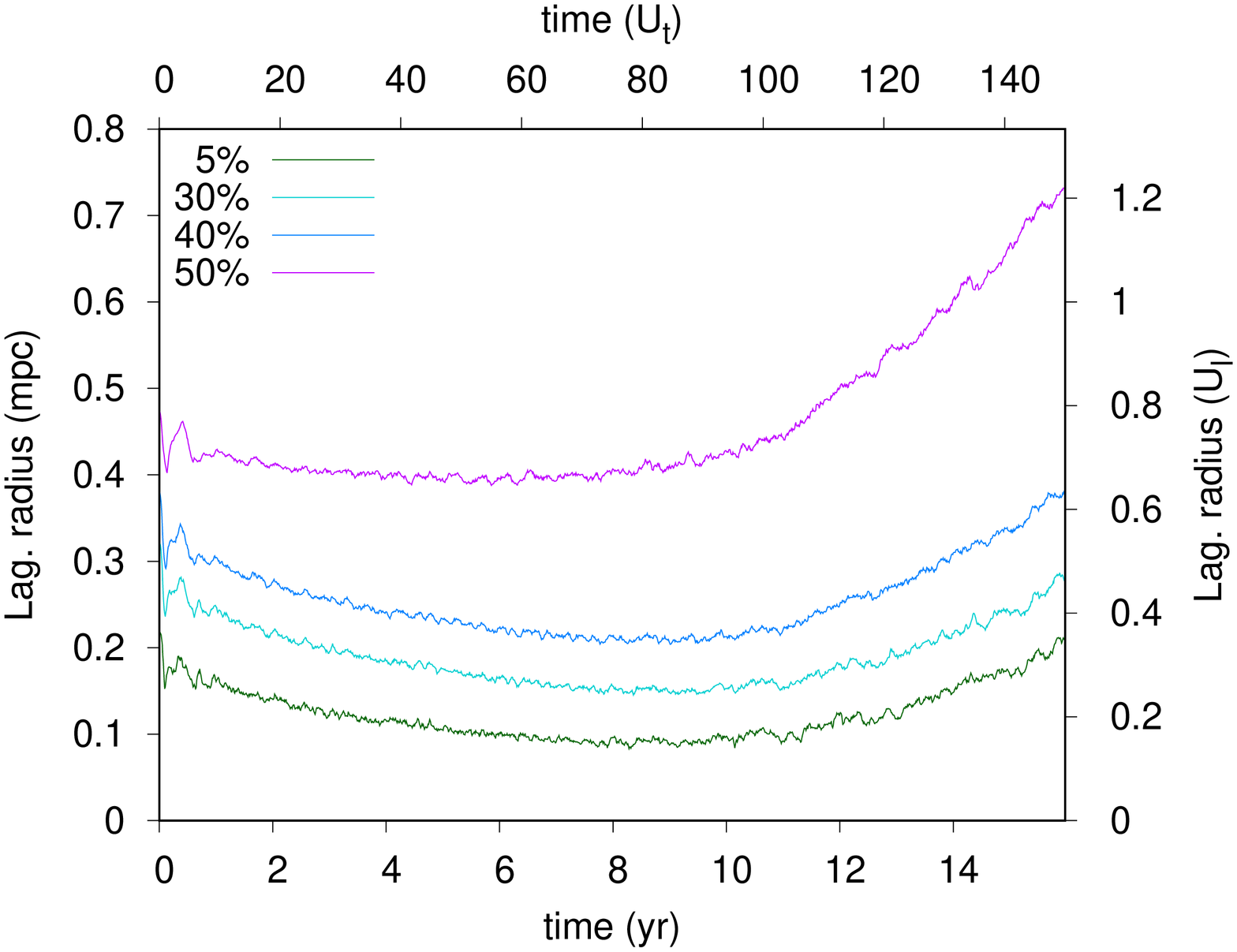} \\
	\includegraphics[width=0.71\columnwidth,angle=-90]{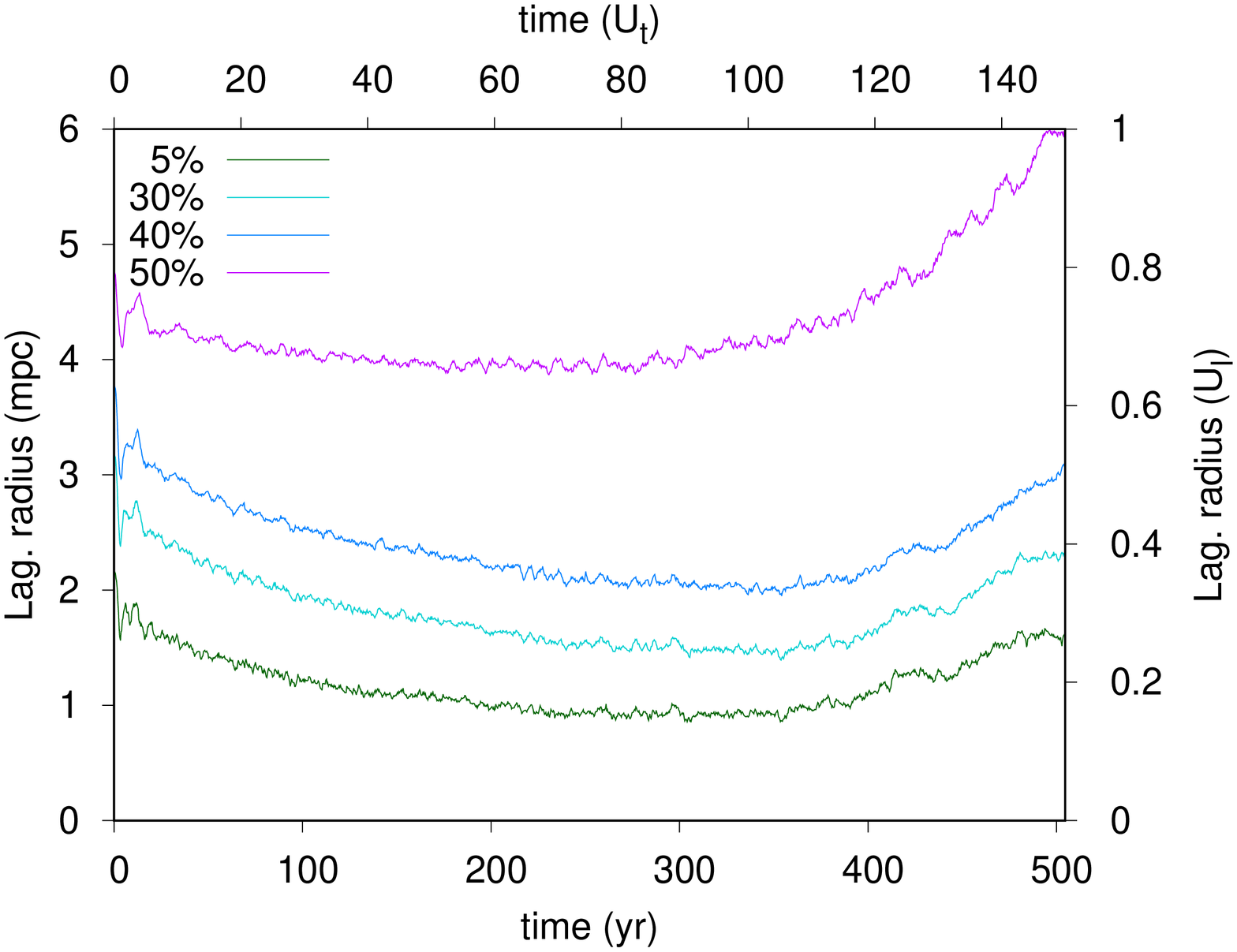}
\end{tabular}
\caption{Initial evolution of the Lagrangian radii (from $5$ percent to $50$ percent) of the system, centered on the center-of-mass of its bound core. Top~: average over the $20$ simulations of set~1. Bottom~: average over the $20$ simulations of set~2.}
\label{fig_lagrangian_radii}
\end{figure}

\begin{figure}
\centering
   \includegraphics[width=0.7\columnwidth,angle=-90]{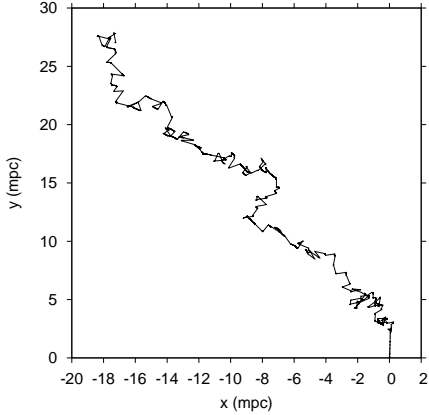}
	\caption{Drift along time of the center-of-mass of the bound core with respect to the center-of-mass of the whole system, for one of the simulations of set~1A.}
    \label{fig_center-of-density}
\end{figure}

Figure~\ref{fig_lagrangian_radii} displays the average (over all the simulations of set~1 and set~2 respectively) evolution of some of the Lagrangian radii of the system.  Note that, due to the non isotropic expulsion of some IMBHs, the center-of-mass of the actual cluster (that is the gravitationally bound part or `core' of the system) deviates  from the position of the center-of-mass of the whole system (see Fig.~\ref{fig_center-of-density}). For a better display we, thus, decided to evaluate the Lagrangian radii with respect to the center-of-mass of the bound core of the system. We defined this bound core of the cluster by excluding those objects which reach with positive energy a distance from the system such as to make very unlikely that they can undergo interactions such to lead them back to negative energy. 

%\textcolor{red}{Update : The distance of exclusion is $300 \textrm{U}_{\subt{l}}$, which means $180$ mpc for set 1 and $1800$ mpc for set 2.}

The Lagrangian radii are evaluated in percentage of the total mass of the cluster and so all the bodies, including possibly escaping IMBHs and growing (in mass) objects, are taken into account. Of course, the escaping IMBHs lead to a natural increase of the high-percentage Lagrangian radii so that in Fig.~\ref{fig_lagrangian_radii} we display only up to the $50$~percent Lagrangian radius ($R_{1/2}$). 

\begin{figure}
\centering
\begin{tabular}{l}
    \includegraphics[width=0.6\columnwidth,angle=-90]{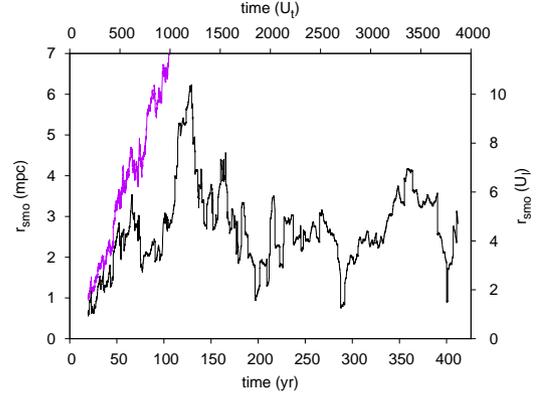}
\end{tabular}
\caption{Black line gives the distance ($r_{smo}$) of the growing super massive object to the center-of-mass of the bound `core' of the system, for one arbitrarily chosen simulation of set~1A, taken as example. For comparison we also give in purple the half mass radius of the system.
The curves are plotted starting from the time when the SMBH "seed" is already formed by 42 merged IMBHs.
 }
\label{fig_distance_smo}
\end{figure}

Note that growing, massive objects, although they remain inside the bound core of the system and close to its center-of-mass, do not a-priori coincide with this center-of-mass (see Fig.~\ref{fig_distance_smo} which shows, also, how the growing BH movement is since the beginning well within the half mass radius) and could, consequently, induce sharp variations in the latter evolution of the low-percentage Lagrangian radii.

\begin{figure}
\centering
\begin{tabular}{cc}
	\includegraphics[width=0.47\columnwidth,angle=-90]{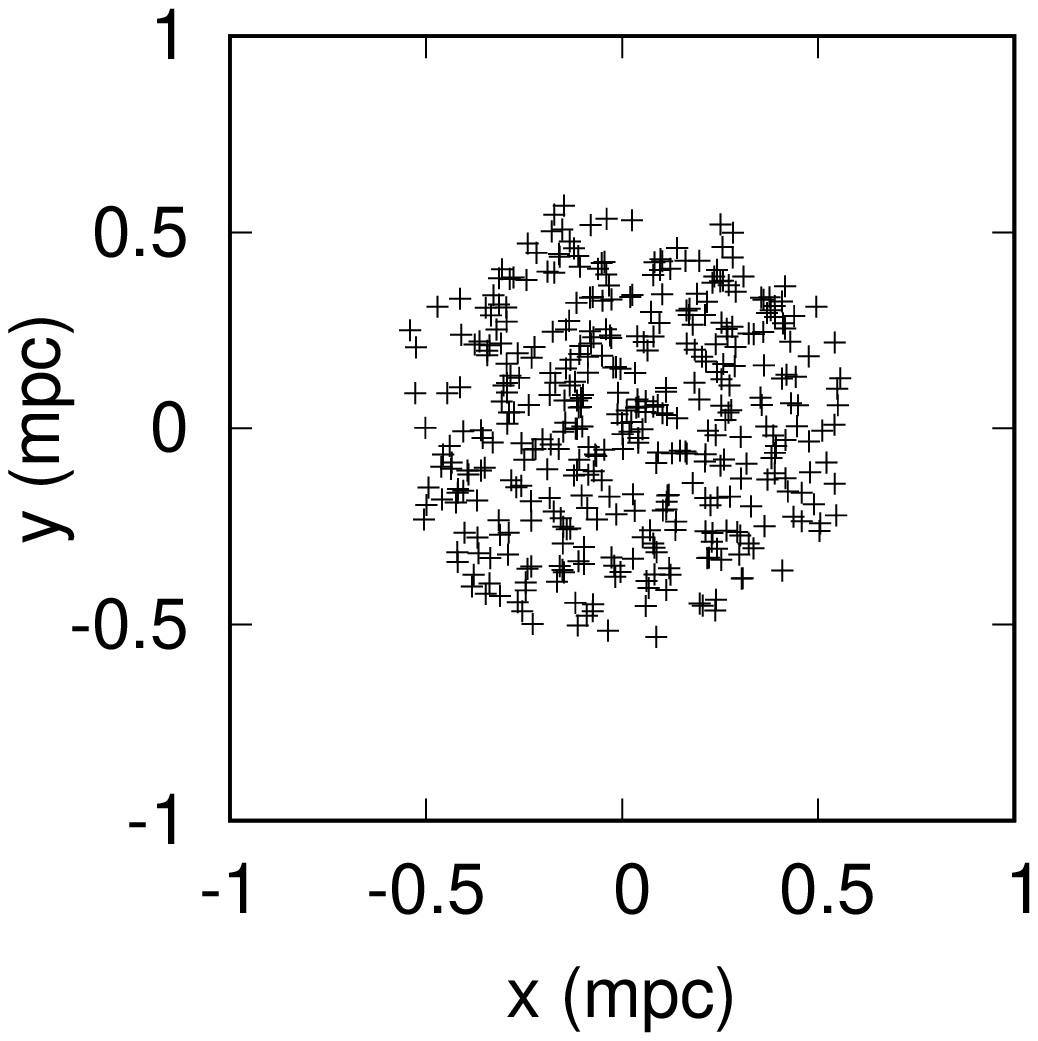} & \includegraphics[width=0.47\columnwidth,angle=-90]{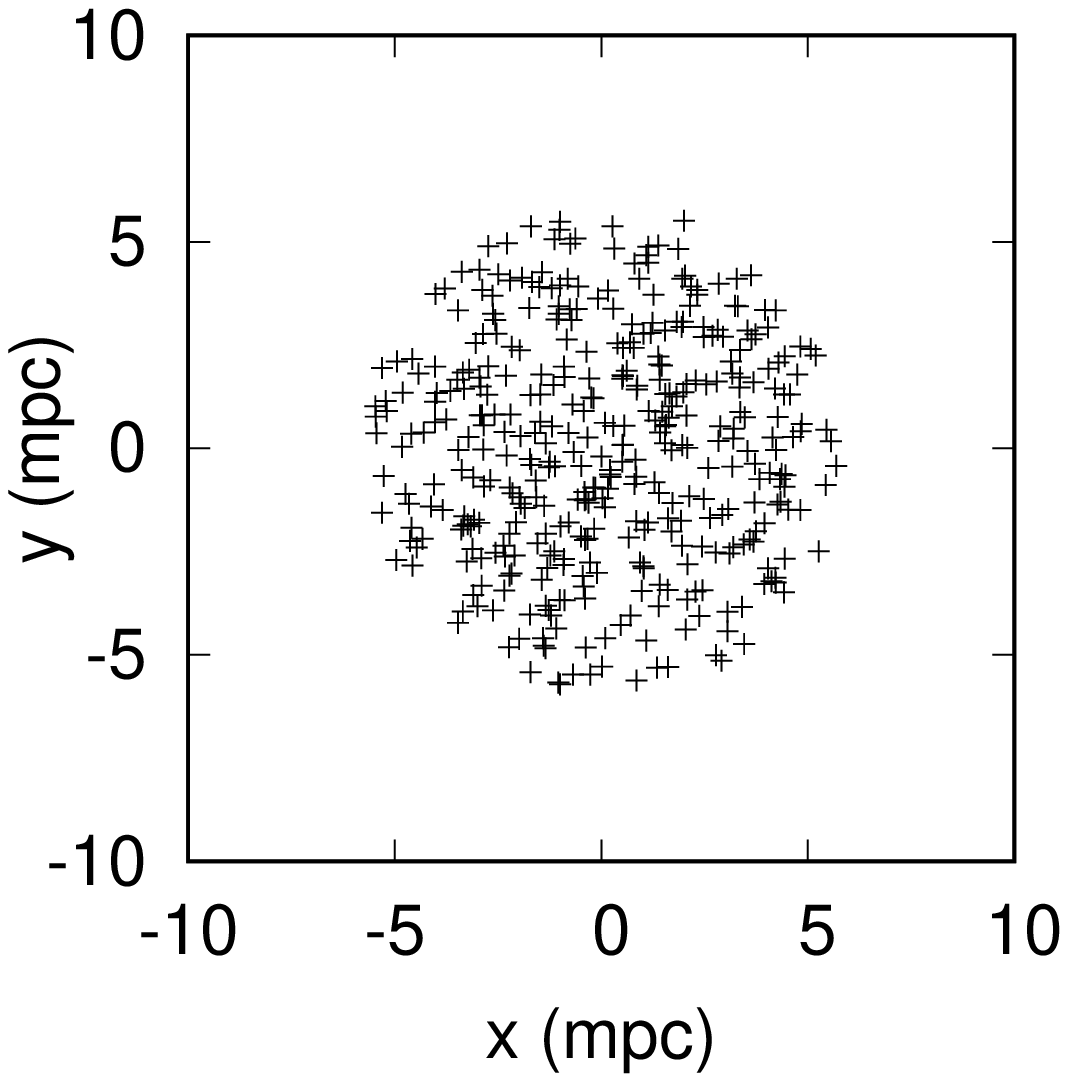} \\
	Set~1A, $t=0$ & Set~2A, $t=0$ \\
	\includegraphics[width=0.47\columnwidth,angle=-90]{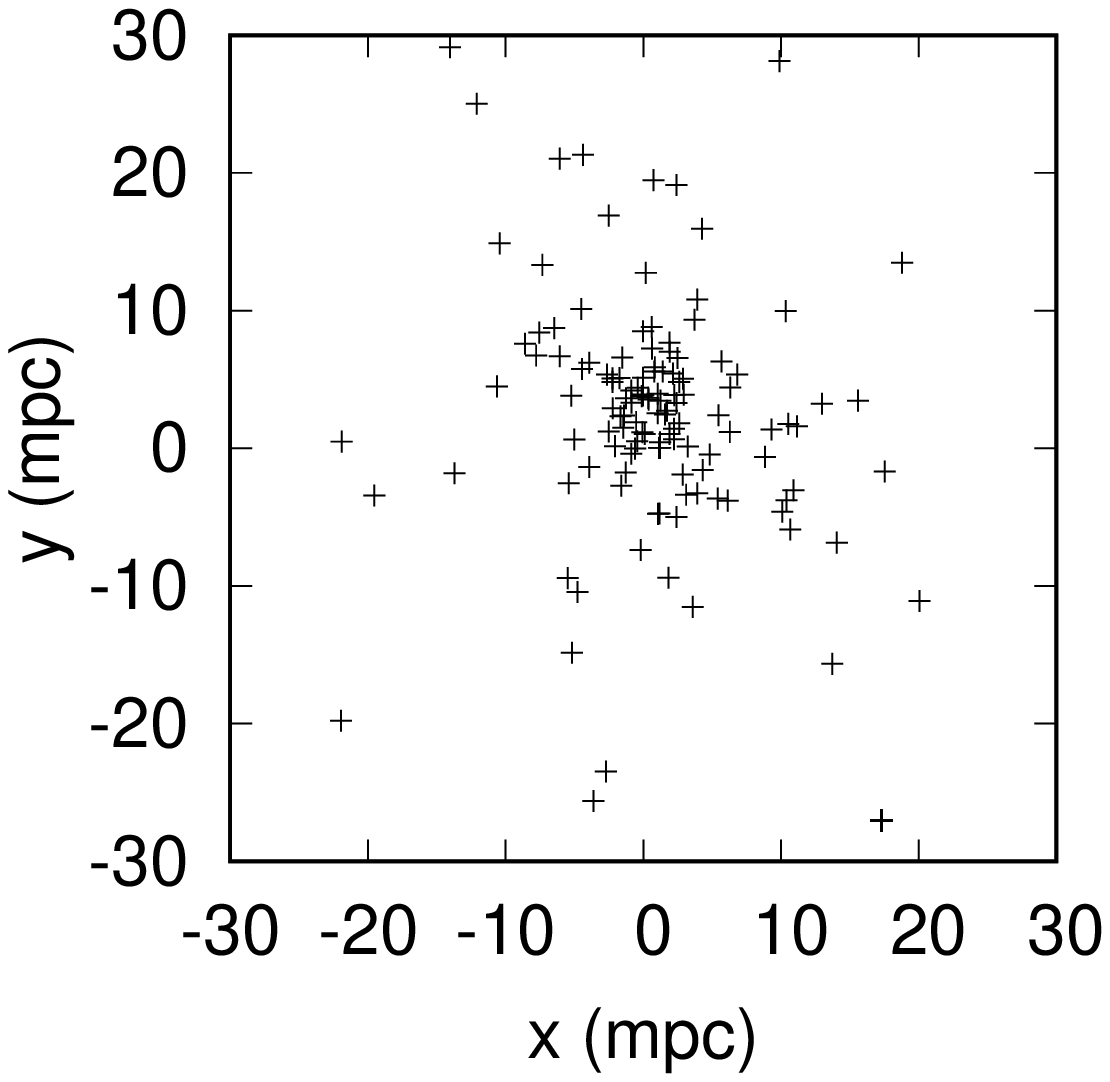} & \includegraphics[width=0.47\columnwidth,angle=-90]{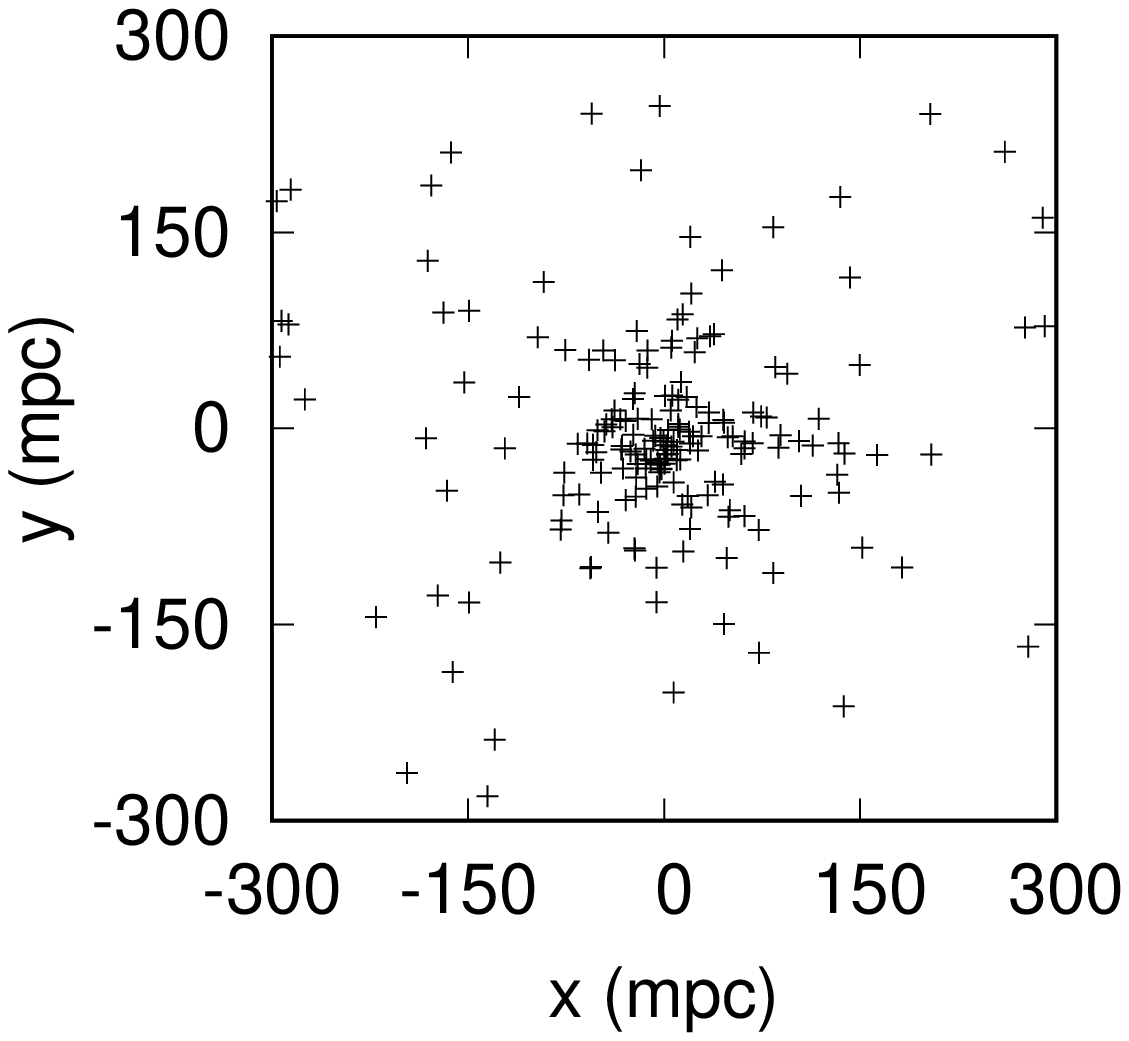} \\
	Set~1A, $t=426$~yr & Set~2A, $t=13\ 455$~yr
\end{tabular}
\caption{Snapshots of the stellar system at $t=0$ and $t= t_{\subt{max}}$ for arbitrarily chosen simulations of set~1A and set~2A, centered on the center-of-mass of the bound core of the system.}
\label{fig_snapshots}
\end{figure}

In Figure~\ref{fig_lagrangian_radii} we see that after a period of contraction lasting, in both cases, about $100$ crossing times, the system expands steadily. The evolution leads to the substantial internal change characteristic of self gravitating systems: an initial homogeneous distribution is remodeled into a dense core surrounded by a low density halo. The snapshots of the system configurations on one of the coordinate planes in Fig.~\ref{fig_snapshots} give a qualitative sketch of this change in the layout of the system. \\

%For this reason, massive objects are usually excluded from the computation of Lagrangian radii. On the other hand, our system does not contains massive objects ab initio, but some of the objects it contains may come to increase in mass through successive merger events. We can not exclude all the merger remnants from the computation, and defining a mass-threshold above which objects are excluded would result in growing massive objects suddenly disappearing from the computation as they reach the threshold. The only reasonable choice is to treat separately the early and latter evolution of our system, which is why Figure~\ref{fig_lagrangian_radii} only reach up to $t = 150 \textrm{U}_{\subt{t}}$. At that time, the heavier object recorded in all of set~1 simulations have aggregated $10.25 \%$ of the total mass of the system, while in set~2 it is only $3.75 \%$.

%\textcolor{red}{If we use this text, we will need to change our computation of the long-term $50\%$ Lagrangian radius. We will have a first phase up to $t = 150 \textrm{U}_{\subt{t}}$ (perhaps $t = 100 \textrm{U}_{\subt{t}}$ for set 1 ?), displayed in Figure~5, then a second phase showing a linear fit for the $50\%$ Lagrangian radius \textbf{excluding} the smo.}

$ $ \\

%\textcolor{roberto}{at the light of what said above, is the following part regarding $R_{1/2}(t)$ still valid?}

%\textcolor{red}{I think it is valid. This is a fit computed including all objects. We could recompute it to exclude the growing supermassive object, but since we said that the growing supermassive object affects only the low-percentage Lagrangian radii, I think it is not necessary. \textbf{But we should add a sentence to justify it, saying that we checked and we are sure that the growing supermassive object stay well within $R_{1/2}(t)$.} I think that, to justify that, we should plot the latter evolution of $R_{1/2}(t)$ together with the distance of the smo to the core of the system on figure 8. This would "prove" our point.}

%\textcolor{roberto}{ok, I wait for this plot (very likely it would be for our use and not necessary to include in the paper. If things do not change much I would not specify that for Fig. 5 we considered the growing massive object and for the $R_{1/2}$ fit not. It would look weird.}

%\textcolor{red}{I updated the plot. So we should keep the computation below as they are, just add a sentence clarifying that $R_{\sub{1/2}}$ is computed the same way as above (and that this is ok because the growing smo only affects only the low-percentage Lagrangian radii ?). We may or may not keep the plotting of $R_{\sub{1/2}}$ on Figure 5, as you prefer (if we keep it, it may be better to change it from purple solid line to black dashed line ?)}

%\textcolor{red}{((Add some comment saying that this fit is valid because the smo is well below $R_{\sub{1/2}}$))}

The late time evolution of the average half-mass radius $R_{\sub{1/2}}$ ($50$~percent Lagrangian radius), which is a good definition of the system radial scale, is well fitted by a linear relation~: 
\begin{equation} 
    R_{\sub{1/2}}(t) \simeq a_{\sub{1/2}} t + b_{\sub{1/2}}.
\label{eq_lagrangian_fitting}
\end{equation}

For set~1 and $t > 15$~yr, the values of the parameters are
\begin{equation}
    \left\lbrace \begin{array}{l}
    a_{\sub{1/2}} = 5.613 \times 10^{-2} \pm 3.10^{-5} \mbox{ mpc yr}^{-1}, \\
    b_{\sub{1/2}} = -0.012 \pm 0.009 \mbox{ mpc},
    \end{array} \right.
\end{equation}
leading to a root mean square error of the fit equal to $0.8$~mpc for $R_{\sub{1/2}}$. For set~2 and $t > 500$~yr, the values are 
\begin{equation}
    \left\lbrace \begin{array}{l}
    a_{\sub{1/2}} = 1.8070 \times 10^{-2} \pm 6.10^{-6} \mbox{ mpc yr}^{-1}, \\
    b_{\sub{1/2}} = -8.09 \pm 5.10^{-2} \mbox{ mpc},
    \end{array} \right.
\end{equation}
giving a root mean square error equal to $5$~mpc. \\

The average half-mass radius at the final simulation time for set~1 is equal to $24$~mpc ($37$~mpc for set~2), that is $50$ times ($8$ times for set~2) the initial half-mass radius, $R_{\sub{1/2}}(0) = 2^{-1/3} R_{\sub{0}}$.

Even with our original safety margin of one order of magnitude for the initial cluster radial size, at the end of the simulation the core of the cluster extends much farther out than the area allowed for our purpose of mimicking the presence of a SMBH by a dense cluster of IMBHs. This is true for every simulation of set~1 and not only on average. This is even more the case for set~2, where no safety margin was taken. This result is not surprising~: we actually expected that the extreme conditions required (a stable system of IMBHs of total mass $M_{\sub{S}} = 4\times 10^6 \,\msun$ and maximal size $\leq 5$~mpc) was very unlikely to be reached. \\

As we see in the next subsection, in both set~1 and set~2 the IMBH cluster undergoes various merger episodes. This has relevant consequences, whose main result is the formation of a very massive BH as coming out from the dominant object growing up after successive merger events.

Hence, we conclude that, as expected on basic theoretical understanding, the answer to the first of the issues we raised in introduction is negative~: a cluster of IMBHs dense enough to mimic the dynamical role of a SMBH would not be stable for a significant time. On the other hand, thanks to our simulations we saw how this instability of the system results in a quick aggregation of mass, efficient enough to beget a super-massive black hole from successive mergers of less massive seeds.

\begin{figure}
\centering
\begin{tabular}{r}
    \includegraphics[width=0.71\columnwidth,angle=-90]{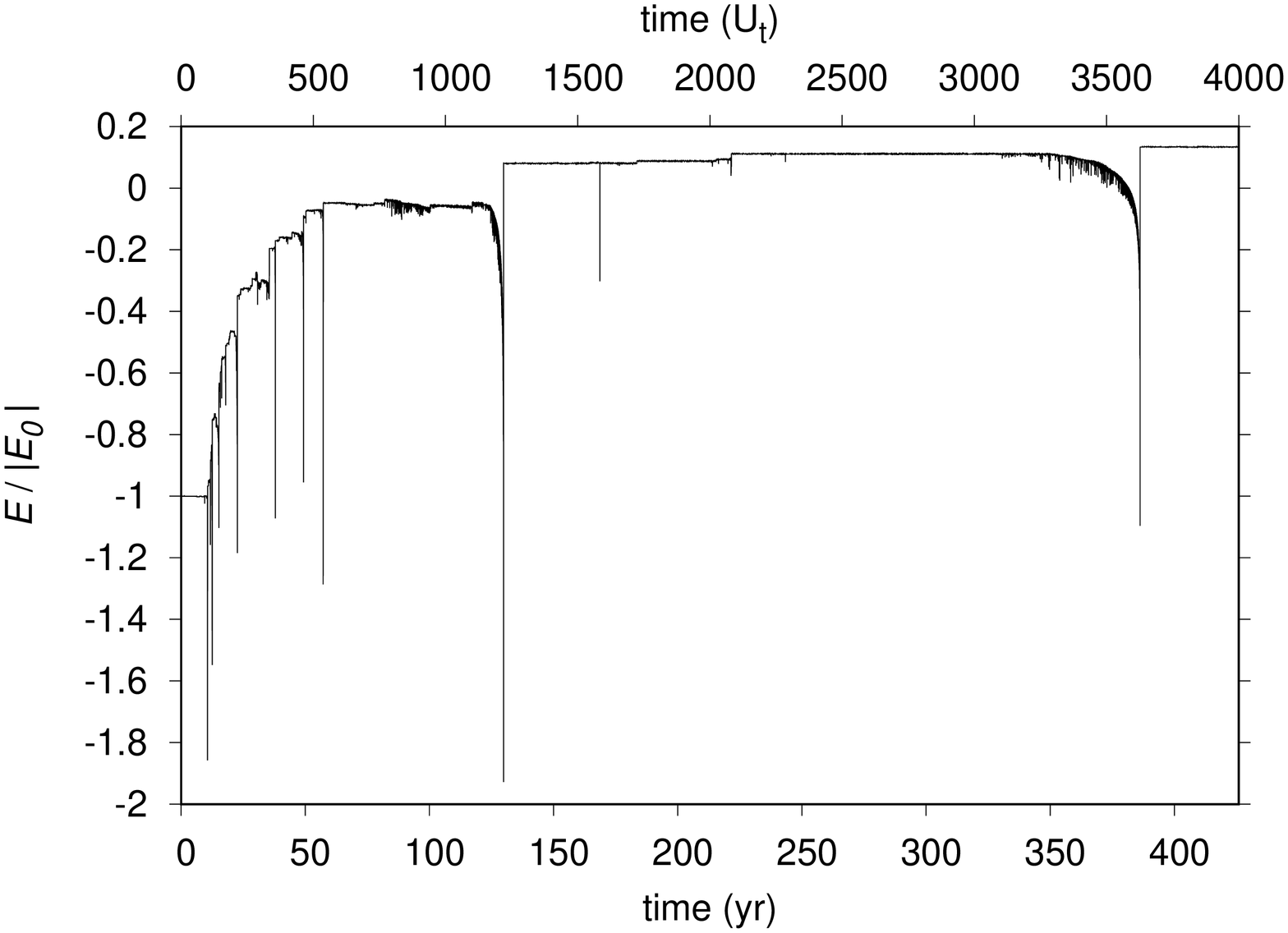} \\
    \includegraphics[width=0.69\columnwidth,angle=-90]{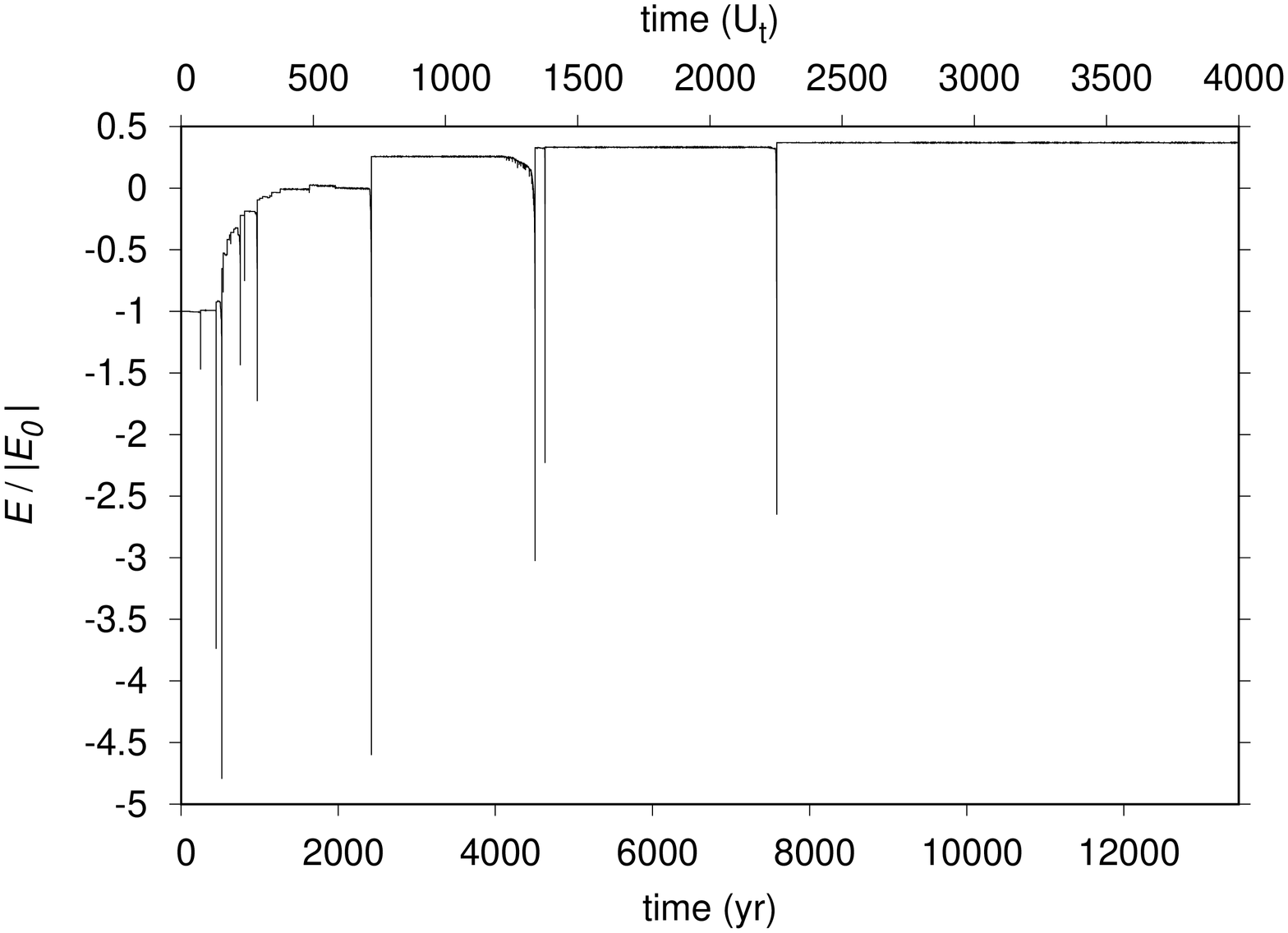}
\end{tabular}
\caption{Evolution of the total mechanical energy $E$ of the system along time, in fraction of its absolute initial value $|E_0|$. Top~: a simulation of set~1A. Bottom~: a simulation of set~2A.}
\label{fig_total_energy_1A01}
\end{figure}

%\begin{figure}
%\centering
%\begin{tabular}{ll}
%    \includegraphics[width=0.3\columnwidth,angle=-90]{Energy_1A01/Energy.eps} &
%    \includegraphics[width=0.3\columnwidth,angle=-90]{Energy_2A10/Energy.eps}  \\
%    \includegraphics[width=0.35\columnwidth,angle=-90]{Mergers/Merging_rate_average_set1.eps} & \includegraphics[width=0.37\columnwidth,angle=-90]{Mergers/Merging_rate_average_set2.eps} \\ 
%    \includegraphics[width=0.3\columnwidth,angle=-90]{Mergers/Merging_percentage_average_set1.eps} & \includegraphics[width=0.3\columnwidth,angle=-90]{Mergers/Merging_percentage_average_set2.eps}
%\end{tabular}
%\caption{Testing the reorganization of plots.}
%\label{fig_test}
%\end{figure}

\subsection{The formation of a super-massive black hole by subsequent mergers}
\label{section_smo}

Along the time evolution of the IMBH cluster under study, many merger events occur. Besides the relevance they have in both growing a super-massive object from the dominant aggregation seed and their repeated bursts of gravitational waves, the merger events have an effect on the overall cluster structure.

Actually, in our simulations the mechanical energy of the system varies due to two phenomena~: one is the energy loss via gravitational radiation (accounted for by the $2.5$ order PN terms in \texttt{ARWV}) during the binary inspiral, while the other is a consequence of collisions, in the way we explained in Sect.~\ref{section_envar}. 

As we saw in Sect~\ref{section_envar}, every merger corresponds to a small injection of positive energy in the system (Eq.~10) so that the total energy of the $N$-body system increases whenever a merger event takes place. As a matter of fact, the total mechanical energy we compute all along our simulations shows a stepwise increase after every merger event (see Fig.~\ref{fig_total_energy_1A01}). After numerous successive merger events, the total energy can eventually become positive, so that the system becomes gravitationally unbound. 

In extremely dense systems such as the ones studied here, close three-body encounters are found to happen often, causing binary pairs to form and tighten, ultimately leading to merger through relativistic final orbital decay. As we mentioned in Sect.~\ref{section_merger_routine}, \texttt{ARWV} is specifically designed to account for these situations, at least until PN approximation maintains its validity. \\

% Because of the comparatively low time-resolution of our simulations, the whole process is sometime too quick to be observed and some merger events seem to happen out of the blue. This is the case for $22$ per cent of the merger events of the first test of set~1A, for example. More than half of these ``instantaneous'' mergers happen in the ten first years of evolution of the system, that is when the density is still very high and three-body encounters all the more common. Of course it is also possible, although extremely unlikely, that one of those mergers arise from a nearly head-on collision. ---- \textbf{add in a sentence to say that the treatment of the collision would still be correct.} \\

\begin{figure}
\centering
\begin{tabular}{r}
    \includegraphics[width=0.66\columnwidth,angle=-90]{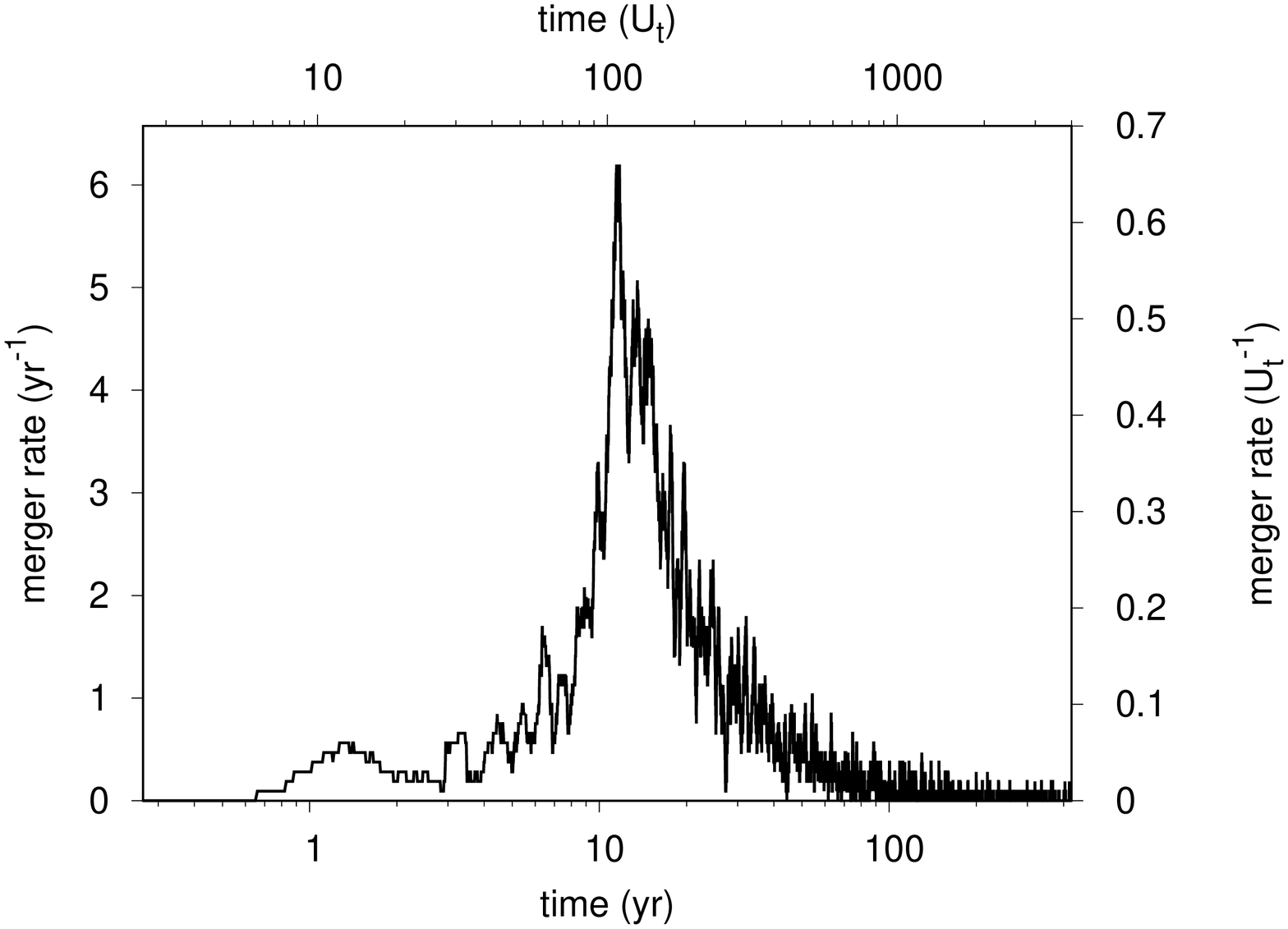} \\
    \includegraphics[width=0.7\columnwidth,angle=-90]{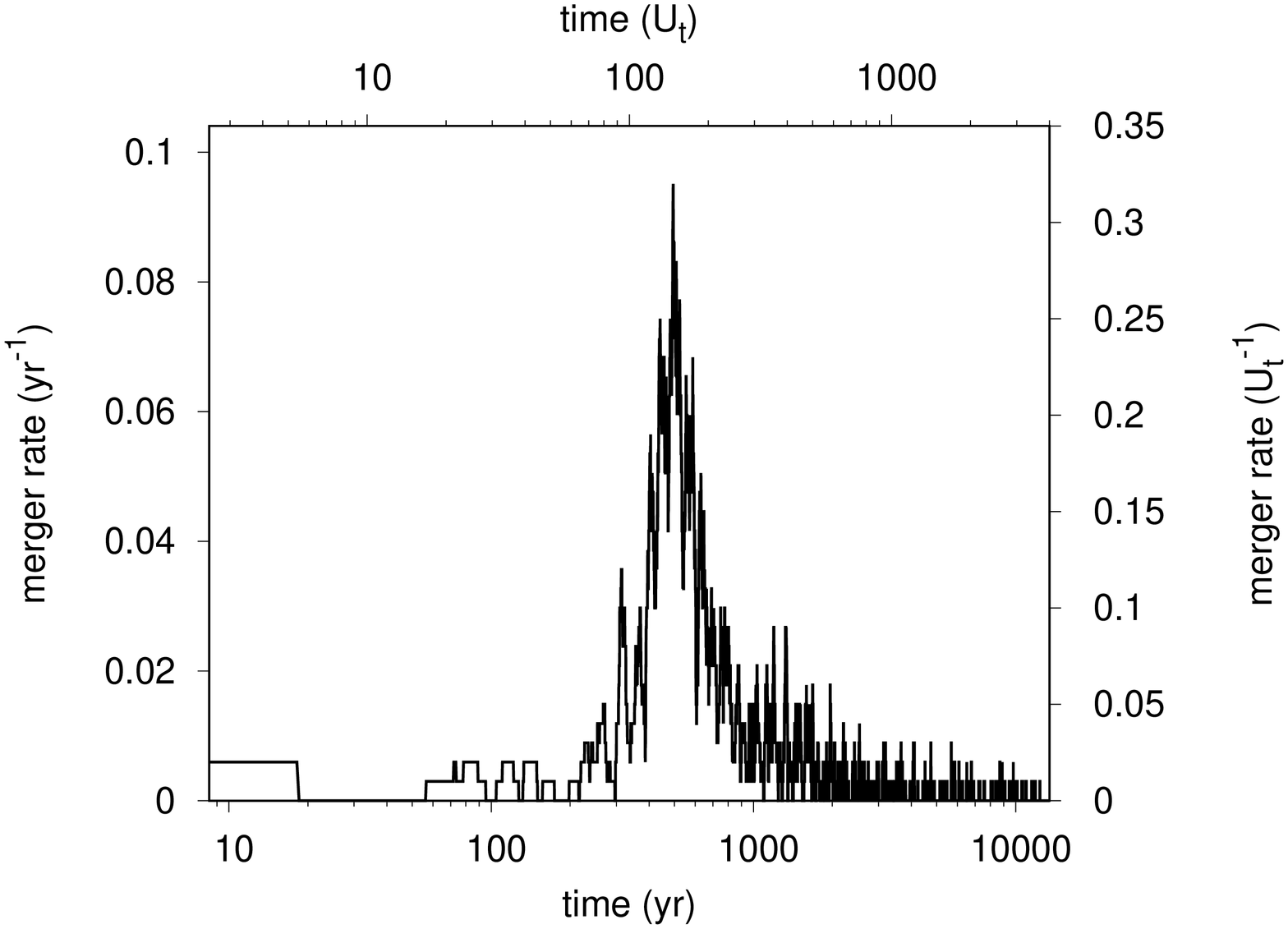}
\end{tabular}
\caption{Top panel~: average rate of merger over the $20$ simulations of set~1. Bottom panel~: average rate of merger over the $20$ simulations of set~2.}
\label{fig_average_merging_rate}
\end{figure}

As Figure~\ref{fig_average_merging_rate} shows, the merger rate comes to a peak at about $100\, \textrm{U}_{\subt{t}}$ after the beginning of the simulation for both set~1 and set~2 (in physical time it is $t \simeq 10$~yr for set~1 and $t \simeq 500$~yr for set~2). This peak time corresponds to the time of maximum compression of the IMBH cluster, as seen in Fig.~\ref{fig_lagrangian_radii}. The merger remnants, if not ejected from the core of the cluster due to high recoil velocity (which, has we have already shown, is a very rare case), constitute an aggregation seed apt to induce further mergers. At this time, the number of merger remnants is maximal (see Table~\ref{tab_merger_remnants} for more details).

\begin{figure}
\centering
\begin{tabular}{c}
    \includegraphics[width=0.7\columnwidth,angle=-90]{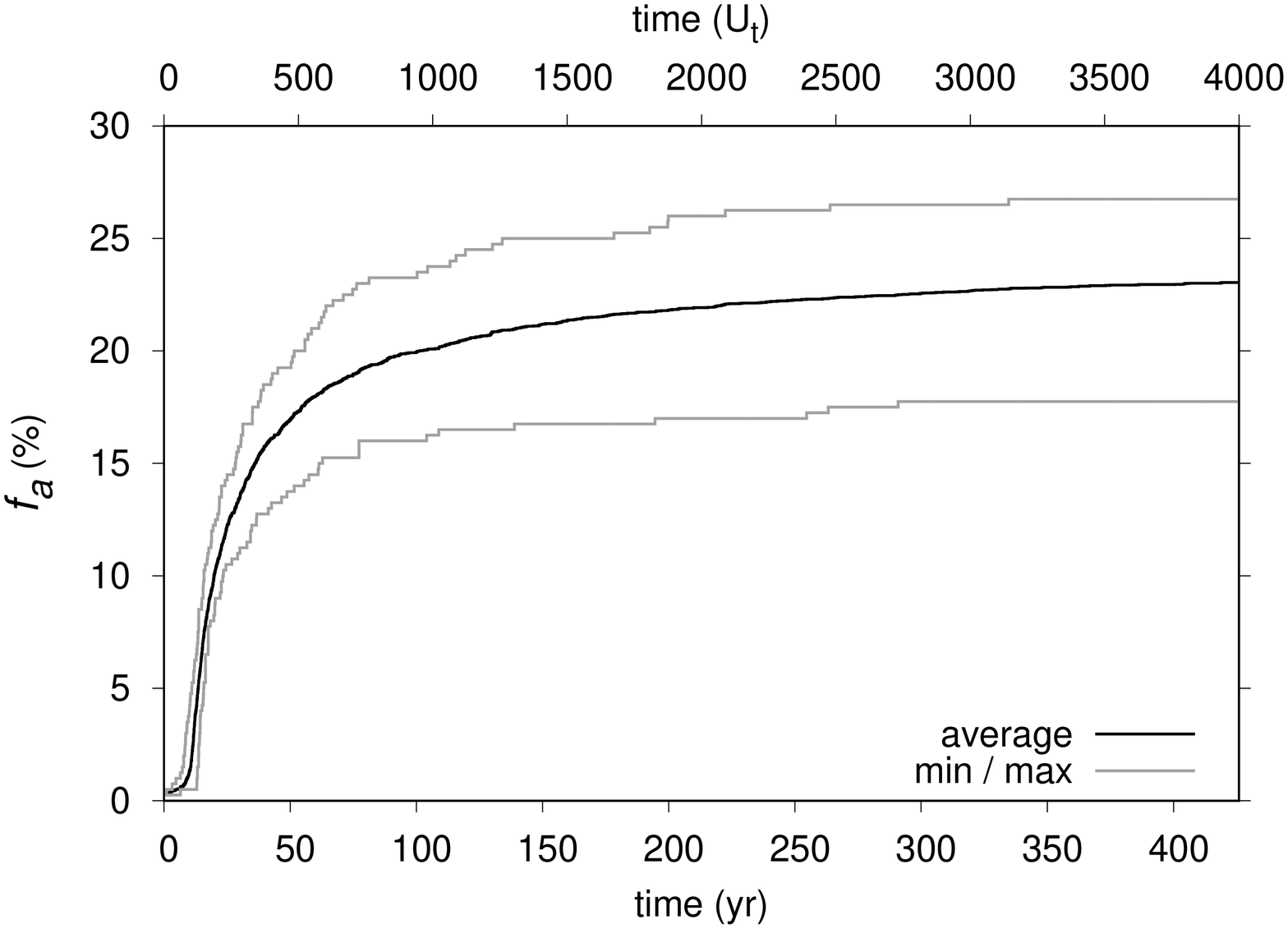} \\
    \includegraphics[width=0.7\columnwidth,angle=-90]{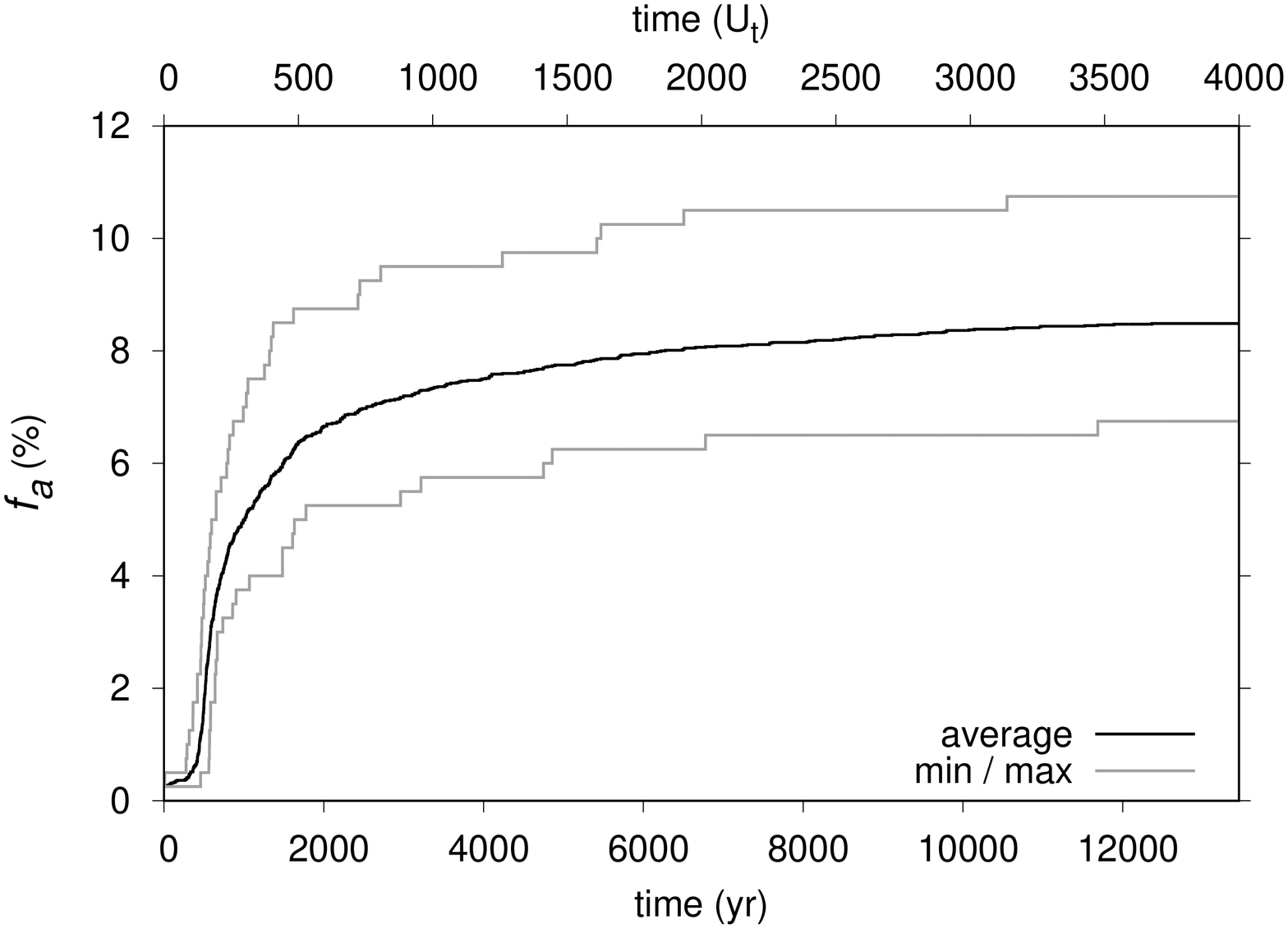}
\end{tabular}
\caption{Percentage, $f_a$, of the stellar mass accumulated into one single body along the simulation. Top~: Average over the $20$ simulations of set~1. Bottom~: Average over the $20$ simulations of set~2.}
\label{fig_merging_percentage}
\end{figure}

\begin{table}
\centering
\begin{tabular}{crlll}
           & Peak time & min.     & max.     & average   \\ \hline 
    Set~1A & $11.45$~yr  & $0.005$ & $0.045$  & $0.0235$ \\ \hline
    Set~1B & $11.04$~yr  & $0.005$ & $0.05$   & $0.0205$ \\ \hline
    Set~2A & $494.46$~yr & $0.01$  & $0.025$  & $0.0175$ \\ \hline
    Set~2B & $493.95$~yr & $0.005$ & $0.0338$ & $0.0139$ \\ \hline
\end{tabular}
\caption{Minimal, maximal and average (over all simulations of the indicated set) fraction of the total mass $M_S$ gone into merger remnants at the time when the merging rate is maximum.}
\label{tab_merger_remnants}
\end{table}
% average over whole set~1 : 11.46yr; 0.005; 0.0525; 0.0245
% average over whole set~2 : 494.3yr; 0.005; 0.035; 0.01575
These remnants merge among themselves rather quickly, leading to a dominant very massive object sitting almost at center of the potential well and ``absorbing'' other bodies. Later, due to the contemporary effects of the expansion of the cluster as a whole and the progressive depletion of IMBHs (many of them having been already captured), the merger rate drops and, so, the mass-aggregation process nearly comes to an end (see Figure~\ref{fig_merging_percentage}).

For set~1, the average number of merger events occurring in $426$ years is $93.05$. On average, only three actual merger remnants survived in the cluster at the end of the simulation, one of which contains almost all the mass aggregated. The mass of this super-massive remnant amounts on average to $23$ per cent of the total initial mass of the system, that is, indeed, $23$ per cent of the mass of the super-massive black hole at the center of the Milky Way. A rough and not completely reliable extrapolation of this result says that an initial number of IMBHs $4.35$ larger (i.e. $1740$ IMBHs of $10^4 \,\msun$ each) would be needed to grow a super-massive black hole of $4 \times 10^6 \,\msun$.

For set~2, the dynamics is less violent, as shown by that the maximal number of contemporary merger remnants is much smaller. During $13\ 455$~yr, an average of $33.9$ merger events happened, leading to the survival of only $2$ remnants, one of which accumulated the mass of $33.95$ initial bodies (that is, $8.5$ per cent of the total mass of the system). 

These are interesting results, because they state the possibility to grow a very massive black hole by the violent interactive dynamics of a set of densely packed intermediate mass black holes. 

We already said that the initial conditions of our system are not the most realistic and that a better modelization (with less extreme hypotheses) should be considered in a further investigation. This new model would likely lead to a smaller rate of mass accretion, as hinted by the fact that set~2, whose initial spatial distribution was extended in radial size for a factor $10$ respect to set~1, shows an approximately $10$ times lesser rate of accretion. But what is really interesting in our present results is that, even if this rate were to decrease by two or even three order of magnitude, a comparable fraction of the BH mass would be aggregated in less than one million years.

\subsection{Gravitational waves from IMBH mergers}
\label{sect_grav_waves}

Let $\dot{E}_{\subt{GW}} \geq 0$ denote the energy radiated away by GW per unit of time (emitted power), so that ${E}_{\subt{GW}}(t) = \int_0^t \dot{E}_{\subt{GW}} \,dt$ is the energy lost by the system from the beginning of the simulation up to time $t$. Proper unit of measure for the energy loss is the absolute value of the initial gravitational (binding) energy $|\Omega_{\sub{0}}|$ of the system, so that we express $\dot{E}_{\subt{GW}}$ either in units of $|\Omega_{\sub{0}}|$~yr$^{-1}$ or in units of $|\Omega_{\sub{0}}|~\textrm{U}_{\subt{t}}^{-1}$.

In set~1, the magnitude of the peak in power emission preceding each merger ranges from $\dot{E}_{\subt{GW}} \sim 10^{-13} \, |\Omega_{\sub{0}}|$~yr$^{-1}$ to $\dot{E}_{\subt{GW}} = 0.6 \, |\Omega_{\sub{0}}|$~yr$^{-1}$, for all $20$ simulations. In set~2, it ranges from $\dot{E}_{\subt{GW}} \sim 10^{-13} \, |\Omega_{\sub{0}}|$~yr$^{-1}$ to $\dot{E}_{\subt{GW}} = 8.3 \, |\Omega_{\sub{0}}|$~yr$^{-1}$, for all $20$ simulations.\\

Most of the merger events occur relatively soon after the beginning of the simulation, when the system is still very dense. Prior to the merger, the two progenitor bodies form a loose binary which is subjected to repeated successive interactions with other IMBHs. At a later stage of its orbital shrinking, the binary starts emitting gravitational waves until it, eventually, merges. 

\begin{figure}
\centering
\begin{tabular}{c}
	\includegraphics[width=0.7\columnwidth]{Mergers/avse_arwv.eps} \\
	\includegraphics[width=0.7\columnwidth]{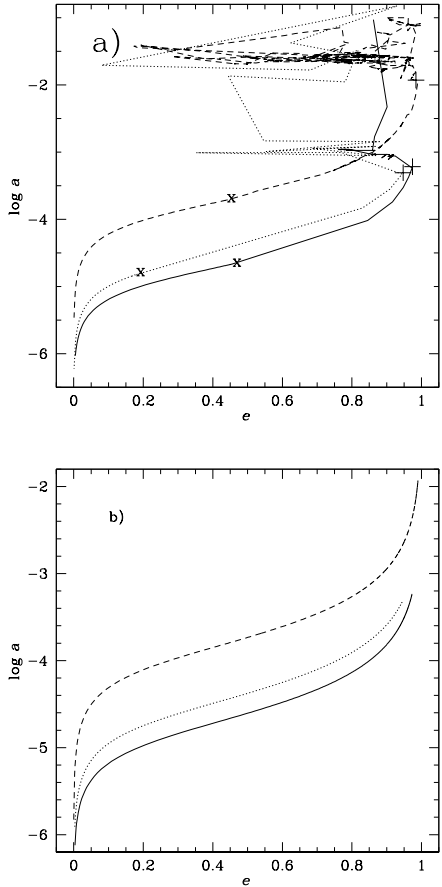}
\end{tabular} 
\caption{Top panel (a): semi-major axis (in units of $\textrm{U}_{\subt{l}}$) vs eccentricity evolution for 3 binaries pertaining to the same simulation of set~2A. The ``+'' symbols mark the beginning of the GW dominated phase. The ``x'' symbols mark the last ARWV output before the merger (see text). Bottom panel (b) : $a$ vs $e$ final evolution according to \citet{pet64} equations, with initial conditions corresponding to the three ``+'' markers in the top panel. Solid line: case 1; dotted line: case 2; dashed line: case 3.}
\label{fig_avse}
\end{figure}

It is notable that while the evolution of binaries formed along the way shows a quite erratic semi-major axis vs eccentricity behavior due to significant external perturbations, when the semi-major axis has shrunk enough (and the eccentricity reached a high value) the final evolution down to the merger resembles, at least for what can be seen by the limited output time resolution of our $N$-body simulations, to that expected in isolation. This is clearly shown in Fig.~\ref{fig_avse}, where the top panel plots $a$ vs $e$ for three sample cases of binaries in set 2A which undergo to a merger. The characteristics of the 3 binary systems undergoing merger are given in Table~\ref{tab_merger_sample}, where the ``initial'' semi-major axis and eccentricity ($a_0$ and $e_0$) are those corresponding to those labeled with a ``+'' symbol in panel a of Fig.~\ref{fig_avse. Actually,} the oscillations in the $a$ vs $e$ relation are caused by passing-by objects perturbations, until (``+'' symbols in  Fig.~\ref{fig_avse} a) the binaries are tight and eccentric enough to evolve independently of the external field. This phase, which leads to the final merger due to GW energy loss, is followed in the ARWV output until the ``x'' symbols. The whole evolution until merging reported in Fig.~\ref{fig_avse} b is obtained, instead, by integration of equations 5.6 and 5.7 in~\citet{pet64}. Notably, the time to merger as obtained by ARWV and by the Peters' like integrations differ by less than $6$ percent. For the sake of clarity and comparison, the bottom panel of Fig.~\ref{fig_avse} gives the $a$ vs $e$ evolution computed by integrating the above mentioned evolutive differential equations from~\citet{pet64}, with initial conditions taken as the ones corresponding to the three ``+'' symbols marked in Fig.~\ref{fig_avse} a).

\begin{table}
\centering
\begin{tabular}{llllllll}
case & $m_1$ & $m_2$ & $q$ & $a_0$ & $e_0$ & $t_m$ & $E_{GW}/(m c^2)$ \\ \hline 
1 & $1$ & $1$ & $1$ & $0.72$ & $0.97$ & $74.21$ & $0.36$ \\ \hline
2 & $2$ & $2$ & $1$ & $0.58$ & $0.94$ & $132.60$ & $0.049$  \\ \hline
3 & $2$ & $29$ & $0.069$ & $14.5$ & $0.99$ & $2252.98$ & $0.021$ \\ \hline
\end{tabular}
\caption{For the 3 cases (as labeled in col. 1: masses in $10^4$ M$_\odot$ (column 2 and 3), mass ratio (col. 4), initial semi-major axis (in AU) and eccentricity (col. 5 and 6), merger time in $\textrm{U}_{\subt{t}}$ (col. 7), and fraction of GW energy released respect to the rest energy ($m=m_1+m_2$) (col. 8). 
}
\label{tab_merger_sample}
\end{table}

In the first case, the two progenitors are basic $10^4 \,\msun$ black holes. In the second, the two progenitors are small merger remnants of mass $2 \times 10^4 \,\msun$. Thus, in both cases the mass ratio is equal to one. These two mergers occur relatively soon after the beginning of the simulation ($t_{m} = 74.21 \ \textrm{U}_{\subt{t}}$ and $t_m = 132.6 \ \textrm{U}_{\subt{t}}$), when the system is still very dense.

In the third case one of the objects in the binary is the growing super-massive black hole and the other a small remnant of a previous merger, giving a mass ratio of $29:2$. This merger occurs later in the simulation ($t_m = 2252.98 \ \textrm{U}_{\subt{t}}$), when the system has largely expanded. The bodies involved in this merger are inside a region where the density is three orders of magnitude less than in the two others considered merger cases. Thus, the encounters with passing-by objects are much less frequent and some phases of the process (loose binary $\rightarrow$ tight binary emitting gravitational wave $\rightarrow$ merger) last longer. In this case the gravitational wave emission phase extends over $16.8$~yr (which is $5$ times longer than the GW emission of the two other merger events displayed in Fig.~\ref{fig_avse} (mainly because of the significantly larger $a_0$ in spite of larger masses and slightly larger $e_0$) with a peak intensity at $2.2 \, |\Omega_{\sub{0}}|$~yr$^{-1}$ and a half-power decay time of $7.8$~days.\\

\begin{figure}
\centering
\begin{tabular}{c}
    \includegraphics[width=0.7\columnwidth,angle=-90]{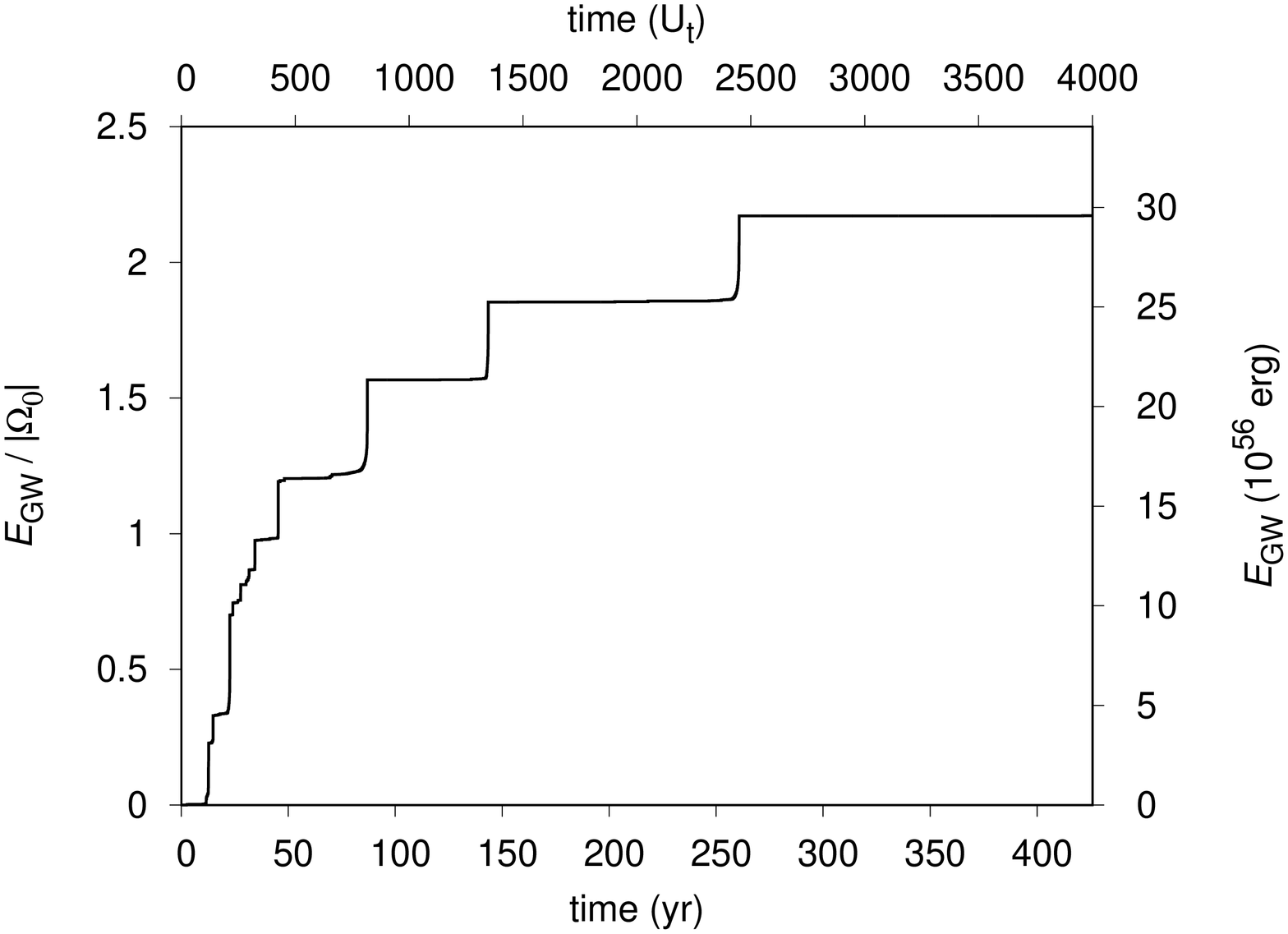} \\
    \includegraphics[width=0.7\columnwidth,angle=-90]{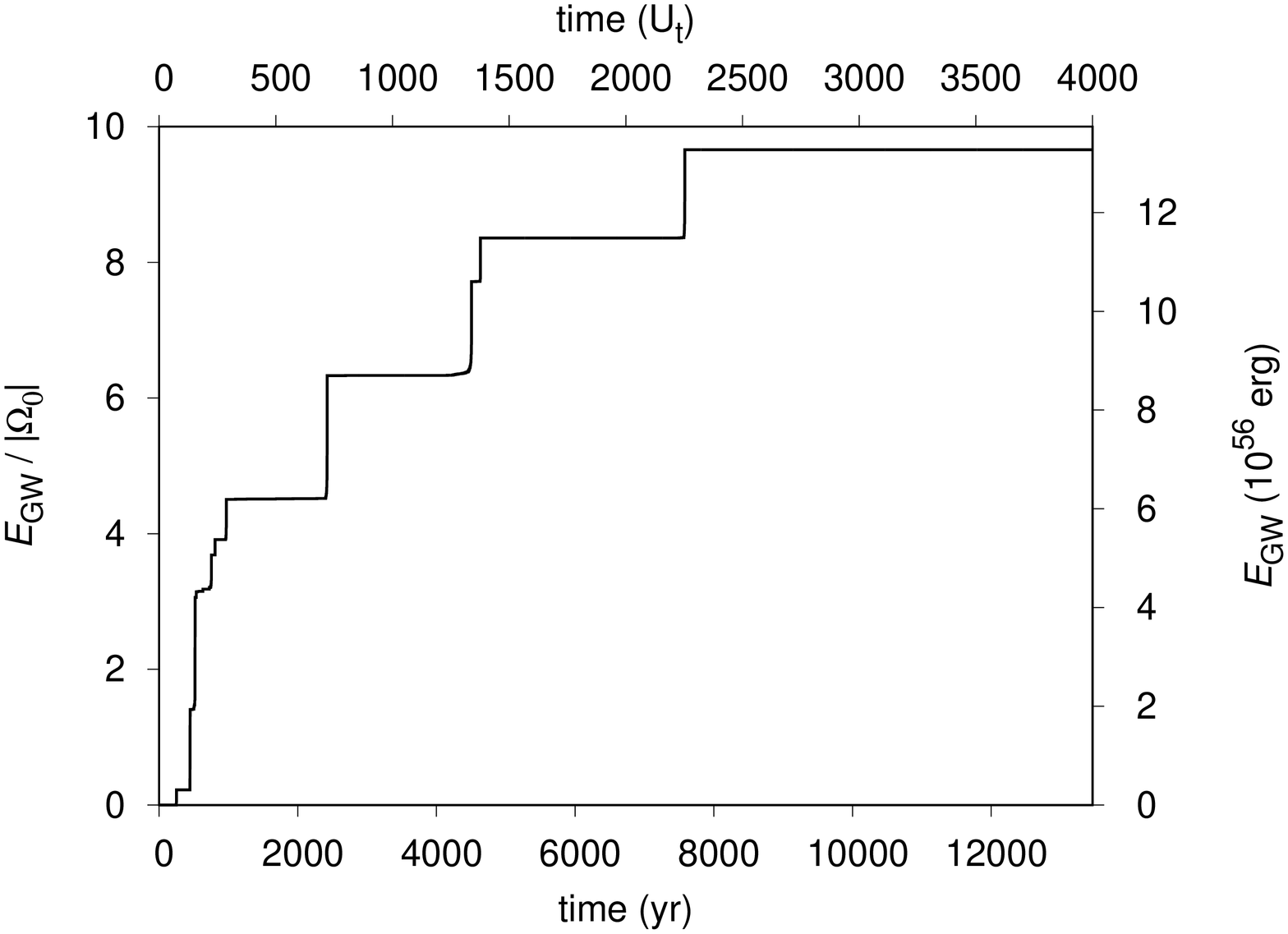}
\end{tabular}
\caption{Energy emitted by the system throughout its evolution, for one of the simulations of set~1A (top) and one of set~2A (bottom).}
\label{fig_grav_energy}
\end{figure}

Figure~\ref{fig_grav_energy} displays an example of the evolution over time of the amount of energy lost by GW, $E_{\subt{GW}}$, in one arbitrary chosen simulation of set~1A (upper panel) and of set~2A (lower panel). The total energy lost by the system at the end of this simulation is equal to $2.56 \, |\Omega_{\sub{0}}|$ for the set~1 case and $9.66 \, |\Omega_{\sub{0}}|$ for the set~2 case. 
Due to the different initial compactness of the two simulated systems, the dynamics of set~1 case is faster, explaining why at the same physical time of $426$~yr (end of set~1's simulations) the GW energy released for set~1 overwhelms that of set~2 case. On the other side, the set~2 case shows a progressive significant GW emission at later times, so that the average time rate of GW emission is not so different in the two cases. If we analyze the output in term of the characteristic time-unit, then in 4000 $\textrm{U}_{\subt{t}}$ the set~2 simulation emits in average 4 times more energy in terms of $\Omega_0$, but 2 times less in absolute value.

For set~1, the average (over the $20$ simulations) quantity of energy lost by the system after $426$~yr is equal to $2.32 \, |\Omega_{\sub{0}}|$. This corresponds to the conversion of $0.082$ per cent of the initial total mass into energy, a little less than $1/3$ of the initial individual IMBH mass.

For set~2, the average (over the $20$ simulations) quantity of energy lost by the system after $13,455$~yr is equal to $6.09 \, |\Omega_{\sub{0}}|$. This corresponds to the conversion of $0.012$ per cent of the initial total mass into energy, a little less than $1/20$ of the initial individual IMBH mass.

\section{Conclusions}
\label{section_conclusion}

In this paper we studied the possible fate of a set of intermediate mass black holes which have allegedly been transported to the Galactic center by their hosting massive star clusters. The possible mechanism of transport and confinement to the central Galactic region has been identified as due to dynamical friction braking of the star background on the motion of massive globular clusters hosting IMBHs. \\

We followed the violent dynamics of this super dense cluster of IMBHs (400 IMBHs of mass $10^4 \,\msun$ each, to give a total mass equal to the one estimated for the Sgr A$^*$ putative black hole) with a high precision $N$-body integrator (\texttt{ARWV}, see \citet{cha19}) containing an accurate treatment of close encounters and of general relativistic effects in the Post Newtonian approximation scheme. We chose two different initial concentrations for the IMBH cluster and included an accurate treatment of recoil velocity after merger following modern GR prescriptions.

Our findings are that~:
\begin{itemize}
    \item the super-dense cluster evolves very fast, without reaching an equilibrium because of the contemporary effect of interactions leading to expulsion of members and the onset of merger events;
    \item the relativistic recoil velocity is rarely high enough to overcome the escape speed, mainly due to that the initial mass ratio, $q$, of the IMBH is $q=1$;
    \item with different efficiency in dependence on the initial number density of the simulated clusters of IMBHs, merger events lead to a dominant ``aggregation'' seed which can grow up in mass to more than $20$ per cent of the initial mass of the cluster;
    \item after this quick growth of what is, actually, a super-massive black hole, the accretion phenomenon slows down due to the dispersal of the residual cluster which makes the further merger cross section exceedingly small;
    \item a simple scaling of our numerical results for the more compact initial cluster considered indicates that a cluster of $1800$ IMBHs with radius $<1$~mpc could lead to the formation of a SMBH of the mass of Sgr A$^*$;
    \item the various mergers, both before and after the onset of a dominant aggregation SMBH seed, generate gravitational waves, whose radiated energy is accounted for by the $2.5$ order terms in the PN approximation. The mergers start as equal-mass merger and proceed toward the regime of IMRIs (intermediate mass ratio inspirals, $m_2/m_1 \sim 100$), and the merging masses are so large that the GW output is peaked at very low frequencies ($< 1$ Hz). The frequency of the emission peak decreases with growing merger mass, such to make them undetectable from ground but still a very appealing source for future space antennas like the joint ESA-NASA satellite interferometer LISA (\url{https://sci.esa.int/web/lisa} and \url{http://lisa.jpl.nasa.gov/}).
    \item the overall evolution of the studied systems, as well the rate of growth of the SMBH is negligible influenced by the individual IMBH spin because of the low value of the recoil velocity after merger with respect to the local escape velocity.
\end{itemize}

This work will be generalized to a more likely framework of IMBHs that are not considered as \textit{ab initio} packed in a narrow region around the Galactic center but that fall progressively there, where they start interacting among themselves.

\section*{Acknowledgements}

We acknowledge support by the Amaldi Research Center (Sapienza, Universit\`a di Roma, I) funded by the MIUR program ``Dipartimento di Eccellenza'' (CUP: B81I18001170001).
We thank Seppo Mikkola for his help in the use and modifications of the \texttt{ARWV} code. 
P. Chassonnery also thanks the Dep. of Physics of Sapienza (Università di Roma, I) for the hospitality during the preparation of this work. A warm thank is also due to R. Schneider for her support during the academic stage of P. C. at Sapienza (Università di Roma, I).\\
Finally, we thank an anonymous referee for specific comments which helped in the presentation of the paper results.
%A particular acknowledge in this context is given by P.C. to prof. R. Schneider.

\section*{Data availability}
The data output of this article will be shared on reasonable request to the corresponding author and is subjected to proper acknwoledgement to this paper.

%\bibliographystyle{mnras}
%\bibliography{bibliography}

\bsp	% typesetting comment
\label{lastpage}

\end{document}